%% file: neurips_2026.tex
\documentclass{article}

\PassOptionsToPackage{numbers,compress}{natbib}

\usepackage[main, final]{neurips_2026}

\usepackage[utf8]{inputenc} 
\usepackage[T1]{fontenc}    
\usepackage{hyperref}       
\usepackage{url}            
\usepackage{booktabs}       
\usepackage{amsfonts}       
\usepackage{nicefrac}       
\usepackage{microtype}      
\usepackage{xcolor}         

\usepackage{amsmath}
\usepackage{multirow}
\usepackage[most]{tcolorbox}
\usepackage{enumitem}

\makeatletter
\renewenvironment{table*}
  {\setlength{\abovecaptionskip}{\@neuripsbelowcaptionskip}%
   \setlength{\belowcaptionskip}{\@neuripsabovecaptionskip}%
   \@dblfloat{table}}
  {\end@dblfloat}
\makeatother

\title{Seeing Beyond the Next Step: World-Model-Guided Human-Like Navigation in Multi-Agent Scenes}

\author{%
  Seong-Eun Hong$^{1*}$ \hspace{1.5em}
  JuYeong Hwang$^{1*}$ \hspace{1.5em}
  Jinhyun Kim$^{1}$ \\[0.3em]
  \bfseries
  Hanyoung Jang$^{2}$ \hspace{1.5em}
  HyeongYeop Kang$^{1\dagger}$ \\[0.5em]
  {\mdseries $^{1}$Korea University \qquad $^{2}$NC AI}
}

\begin{document}

\maketitle

{\renewcommand{\thefootnote}{}%
 \footnotetext{%
  \begin{tabular}[t]{@{}l@{\hspace{0.6em}}l@{}}
    \multicolumn{2}{@{}l}{$^{*}$Equal contribution. \quad $^{\dagger}$Corresponding author.} \\[0.15em]
    Emails: &
    \begin{tabular}[t]{@{}l@{\hspace{1.5em}}l@{\hspace{1.5em}}l@{}}
      \texttt{seong\_eun@korea.ac.kr} & \texttt{05judy02@korea.ac.kr} & \texttt{rlawlsgus0623@korea.ac.kr} \\
      \texttt{ncdrjang@ncsoft.com}    & \texttt{siamiz\_hkang@korea.ac.kr} &
    \end{tabular}
  \end{tabular}}}
 
\begin{abstract}
Selecting a next motion in a populated scene is not just a matter of avoiding what is currently in the way: an action can be locally feasible yet drift onto the road instead of the crosswalk, cut into a stream of crossing pedestrians, or bottleneck a narrow passage. These failure modes are invisible from the current state alone, and acting on them requires reasoning about what each candidate action will cause. World models offer exactly this, namely action-conditioned predictions of the future, but rolling them out at every decision step is a cost deployable navigation systems cannot pay, and the field has largely traded foresight for speed. We argue this is a false dichotomy. The natural place for a world model in human-aware navigation is at training time, as a preference oracle that distills foresight into a fast policy, rather than at inference time, as an online planner. We pair a discrete visual action prior with two task-aligned world models, one for nearby social dynamics and one for future scene semantics, that score candidate actions by their predicted consequences and refine the policy through group-relative preference distillation. At deployment, both world models are removed and the policy acts in a single forward pass. Across five held-out evaluation scenarios, this yields navigation that is simultaneously safer, more socially and group-coherent, and more goal-directed than strong reactive and learned baselines, with naive observers rating the resulting motion as more human-like. Taken together, our findings point to a different approach for bringing world-model foresight into deployable agents: not accelerating rollouts, but amortizing them into the policy so that the cost of looking ahead is confined to training. Project page: \url{https://seoneun.github.io/-NeurIPS-2026-Seeing-Beyond-the-Next-Step/}.
\end{abstract}

\section{Introduction}

An agent can clear every person, vehicle, and obstacle around it at the current moment and still, two seconds later, end up cutting through a group already moving as one, drifting onto a stretch of road a turning car is about to reach, or blocking a doorway someone is about to step out of --- each of these is locally feasible, geometrically valid, and consistent with a reasonable demonstration.
None of them is good behavior. The shared property of these failure modes is that they are invisible from the current state alone: they are determined by what the chosen action will cause, not by what it currently avoids.

Acting well in such scenes therefore requires foresight, anticipating how nearby people, dynamic objects, and the navigable layout will evolve under each candidate motion. But foresight is expensive, and the field has mostly traded it away for speed.
Geometric reactive controllers decide from instantaneous positions and velocities~\citep{helbing1995social,van2011reciprocal}, and learned social and reinforcement-learning policies fold longer-horizon trade-offs into a value function but still act on the current state at decision time~\citep{panayiotou2022ccp,charalambous2023greil,panayiotou2025cedrl}. 
Vision-based steering methods and recent vision-language-action (VLA) models extend this paradigm to richer visual and semantic perception~\citep{dutra2017gradient,brohan2022rt1,pmlr-v229-zitkovich23a,kim2024openvla}, but they invest that capacity in reading the present scene more carefully, not in asking what each candidate action would actually produce. 
World models can ask exactly that question through action-conditioned rollouts \citep{hafner2019planet,hafner2020dreamer}, but they have largely been built as monolithic generators of photorealistic or latent-state futures.
Rolling one out at every step, over a vocabulary of roughly a hundred plausible motions, is a cost the live decision loops of crowd simulation, embodied interactive agents, and on-robot navigation cannot afford. As a result, world models remain largely unexplored in human-aware navigation: the field has effectively been forced to choose between foresight and deployability, and with it, human-like behavior --- by which we mean motion that is perceptually preferred over baselines by naive observers and aligns with demonstrated human trajectories along measurable axes.

We argue that this is a false dichotomy, and that the natural place for a world model in human-aware navigation is at training time, as a preference oracle, rather than at inference time, as an online planner. We instantiate this view in a human-aware navigation policy through three deliberate design choices. First, we train a VLA-style visual action prior over a discrete vocabulary of short-horizon motion anchors, so that candidate actions already lie on a behaviorally plausible manifold shaped by demonstration. Second, we replace the usual monolithic, photorealistic world model with two task-aligned predictors. The social branch predicts relative position and velocity with a direction term capturing who is heading toward whom. The scene branch predicts only the navigation-relevant slice of the future scene, with pedestrians excluded so the two predictors specialize cleanly on motion and layout.
Third, we keep both world models entirely on the training side: their predictions are converted into a compact preference signal and distilled into the policy via Group Relative Policy Optimization (GRPO). The resulting policy inherits the look-ahead reasoning of the world models without paying their cost --- at deployment, it selects an anchor with a single forward pass, and the world models are not needed.

Across five held-out evaluation scenarios, this design outperforms all baselines in goal success and social-region violation on every scenario, and achieves the lowest or tied-lowest group distance on four of the five, with violation rates reduced by up to an order of magnitude without sacrificing goal completion. A user study with naive observers further confirms that our policy's rollouts are perceived as significantly more human-like than those of all three baselines, showing that world-model-based foresight can be brought into human-aware navigation and yield human-like behavior without sacrificing real-time deployability.

Our contributions are:
\begin{itemize}[leftmargin=1.5em,labelsep=0.5em]
    \item \textbf{World models as training-time preference oracles for human-aware navigation.} 
    We frame action selection as ranking locally plausible candidates by their predicted consequences, and realize this through a visual action prior refined by GRPO-based preference distillation, yielding a policy that acts in a single forward pass with no world-model rollout at inference.
    \item \textbf{Dual task-aligned world models, not a monolithic generator.} Our key insight is that future prediction for human-aware navigation decomposes naturally along two axes, social dynamics and scene layout, and we instantiate each branch with a navigation-aligned target: position-velocity with a heading-aware loss, and pedestrian-excluded scene semantics. Ablations confirm that the two branches catch complementary failure modes.
   \item \textbf{Train-time foresight, test-time efficiency, and human-like rollouts.} With world models used only as training-time preference oracles, the resulting policy outperforms all baselines in goal success and social-region violation on all five held-out evaluation scenarios, achieves the lowest ADE against demonstrated human trajectories, and is judged significantly more human-like by naive observers.
\end{itemize}

\section{Related Work}

\paragraph{Human-like navigation and crowd simulation.}
Human-like navigation in this work is treated as a crowd-behavior problem: an agent must move in a way that is safe, socially plausible, and consistent with the local pedestrian flow. Human locomotion is anticipatory rather than purely reactive, and crowd self-organization often emerges from mutual prediction among nearby people \cite{wolpert1995internal,higuchi2013visuomotor,murakami2021mutual}. This motivates our central question: when several motions are locally feasible, which one will remain human-like after the scene evolves?

Classical crowd models and vision-based steering methods provide efficient local interaction rules from current geometry or visual evidence \cite{helbing1995social,van2011reciprocal,dutra2017gradient}. Data-driven and learning-based crowd simulators further improve realism by learning from demonstrations, reference trajectories, or reinforcement-learning objectives \cite{lerner2007crowds,lee2007group,Charalambous2014PAGCrowd,Zhao2017CLUST,panayiotou2022ccp,charalambous2023greil,panayiotou2025cedrl}. These works are the closest baseline family for our experiments because they target plausible crowd behavior. However, most of them either react to the current state or encode future effects implicitly inside a policy or reward. Our work instead makes the future consequences of candidate motions explicit during training.
Other approaches plan with model predictive control over the outputs of trajectory forecasters~\cite{poddar2023crowd} or use conformal prediction to account for forecast uncertainty~\cite{yao2025towards}.

\paragraph{VLA models and anchor-based action priors.}
Recent vision-language-action models show that strong visual-semantic backbones can be adapted into policies that map rich observations to actions \cite{brohan2022rt1,pmlr-v229-zitkovich23a,kim2024openvla}. We use this idea as a visual action prior for crowd navigation, not as an end-to-end solution. In crowded scenes, the difficulty is often not proposing one plausible action, but choosing among multiple plausible actions whose downstream effects differ.

To make this comparison tractable, we represent short-horizon motion with trajectory anchors. Anchor-based prediction has been effective for multimodal future modeling because a compact set of trajectories can approximate diverse local motion modes \cite{chai2019multipath,phanminh2019covernet}.
Low-rank and anchor-based descriptors likewise approximate pedestrian trajectories compactly~\cite{bae2023eigentrajectory,bae2024singulartrajectory,shi2023trajectory}.
Pedestrian forecasting work similarly shows that crowd motion is multimodal and shaped by social and scene context \cite{alahi2016sociallstm,gupta2018socialgan,sadeghian2019sophie,salzmann2020trajectron,yuan2021agentformer,bae2024lmtraj}. Our anchors differ from forecasting outputs: they are executable candidate motions for the controlled agent. This design gives the policy a human-like action manifold from demonstrations, while leaving the final choice to future-aware preference learning.

\paragraph{World models for consequence-aware policy refinement.}
World models are attractive because they allow actions to be evaluated by their predicted consequences rather than only by the current observation. In model-based learning, latent dynamics models such as PlaNet and Dreamer use imagined rollouts for planning or policy optimization \cite{hafner2019planet,hafner2020dreamer}. More recent visual world models scale this idea toward interactive simulation, controllable visual generation, driving futures, navigation rollouts, and game-like environments \cite{yang2024unisim,bruce2024genie,hu2023gaia,gao2024vista,bar2025navigation,guo2025mineworld,sun2025worldplay,tang2025hunyuan,li2026wildworld}. Their main goal is usually to build a simulator, planner, data generator, or future evaluator.

Our contribution is different in the role, representation, and deployment use of the world model. First, the world models are used only during training as a preference oracle; they are removed at inference, so the final policy selects an anchor with a single forward pass. Second, we avoid a monolithic photorealistic future generator. Instead, we use two task-aligned predictors: a State Transition World Model for future relative position--velocity dynamics, and a Semantic Transition World Model for future navigation-oriented segmentation. This follows the idea that decision-relevant semantic futures can be more useful than raw RGB prediction for downstream action selection~\cite{luc2017predicting}. Third, the world models do not directly control the agent. They score candidate anchors by goal progress, smoothness, interaction risk, and semantic validity, and these relative preferences are distilled into the VLA-style anchor policy through GRPO \cite{shao2024deepseekmath}.

\begin{figure*}[t]
\centering
\includegraphics[width=\textwidth]{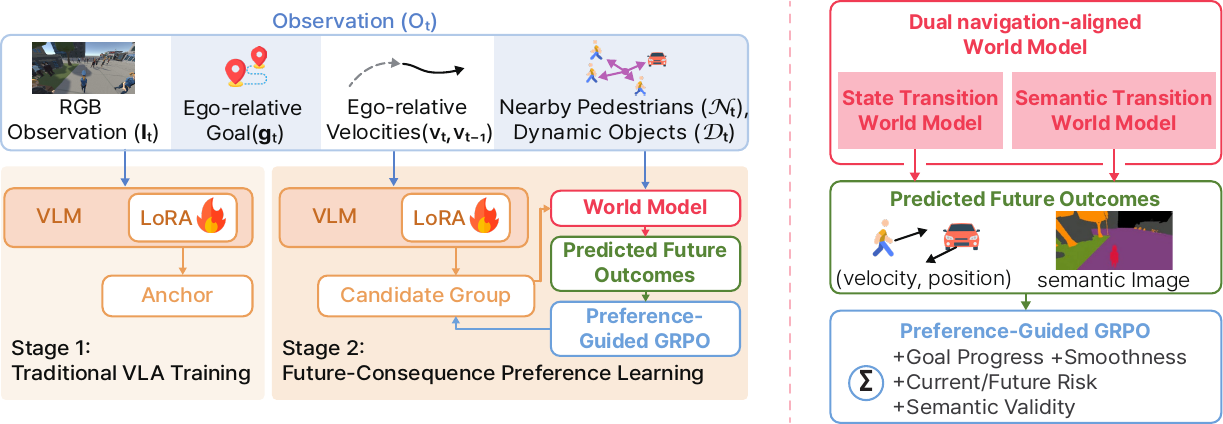}
\caption{\textbf{Overview of the proposed training pipeline.}
In Stage 1, we adapt a pretrained VLM into a VLA policy through supervised learning, grounding agent-centric visual observations and structured navigation context to executable motion-skill actions.
In Stage 2, we further refine the policy with dual world models, which teaches the agent action--outcome knowledge by comparing plausible candidate actions according to their predicted social, semantic, and goal-directed consequences.}
\label{fig:overview}
\end{figure*}

\section{Method}


Our pipeline learns a deployable policy in two stages, as shown in~\autoref{fig:overview}. We first train a \textit{visual action prior} that imitates demonstrated motion over a discrete vocabulary of human-like motion anchors. We then refine this prior with two navigation-aligned world models that score each candidate anchor by its predicted social and semantic consequences, and distill the resulting preferences back into the policy via Group Relative Policy Optimization. Crucially, the world models are used only during refinement: at deployment, the refined policy selects an anchor with a single forward pass, and no world-model rollout is performed.

We use the following notation throughout. At simulation step $t$, the controlled agent receives the observation
\begin{equation}
o_t = \{I_t, g_t, v_t, v_{t-1}, \mathcal{N}_t, \mathcal{D}_t\},
\end{equation}
where $I_t$ is the RGB observation, $g_t\in\mathbb{R}^2$ is the ego-relative goal, $v_t,v_{t-1}\in\mathbb{R}^2$ are the current and previous ego velocities, and $\mathcal{N}_t$, $\mathcal{D}_t$ are the sets of nearby pedestrians and dynamic objects. Each surrounding entity $j$ stores an ego-frame relative position $p_{t,j}\in\mathbb{R}^2$ and relative velocity $u_{t,j}\in\mathbb{R}^2$. The policy is a categorical distribution $\pi_\theta(a_k \mid o_t)$ over an anchor vocabulary $\mathcal{A} = \{a_1, a_2, \dots, a_K\}$, where each anchor $a_k = \{y_{k,0}, y_{k,1}, \dots, y_{k,L}\}$ is a short ego-relative trajectory and $y_{k,\ell}\in\mathbb{R}^2$ is the displacement at anchor step $\ell$.

\subsection{Action Prior over Motion Anchors}
We start from a discrete vocabulary because, over a short horizon, the space of plausible human motion is well approximated by a manageable set of motion modes. This turns action selection into classification over a finite, behaviorally meaningful vocabulary, which (i) anchors candidate actions on a human-like manifold from the outset, and (ii) makes consequence-based comparison tractable.

\paragraph{Anchor construction.}
We build $\mathcal{A}$ offline from training trajectories, recorded at 25 FPS. Each demonstration is cut into 20-frame clips with a 10-frame stride, yielding ego-relative displacement sequences $x_n = \left[x_{n,1},y_{n,1},\ldots,x_{n,20},y_{n,20} \right]\in\mathbb{R}^{40}$. We run $K$-means with $K=100$ in this 40-dimensional space; each cluster center is reshaped into 20 two-dimensional points to form one anchor.  We discuss the choice of $K$ and show nearest-anchor reconstructions in~\autoref{app:anchor_vocabulary}.

\paragraph{Supervised training.}
The first stage trains $\pi_\theta$ as an anchor classifier. We construct a supervised dataset $\mathcal{D}_{\mathrm{SFT}}$, where each observation $o_t$ is paired with the anchor $a_t^\star\in\mathcal{A}$ closest to the demonstrated future trajectory at time $t$. We minimize
\begin{equation}
\mathcal{L}_{\mathrm{SFT}}
= - \mathbb{E}_{(o_t,a_t^\star)\sim \mathcal{D}_{\mathrm{SFT}}}
\left[
\log \pi_\theta(a_t^\star \mid o_t)
\right].
\end{equation}
We instantiate $\pi_\theta$ with a pretrained multimodal backbone and adapt it with LoRA~\citep{hu2022lora}. 
This yields a strong human-like imitation prior, but imitation alone cannot distinguish locally plausible actions with different downstream consequences.

\subsection{Navigation-Aligned World Models}
Refining the action prior requires asking, for each candidate anchor, \textit{what would happen if it were executed?} A natural choice would be to train a single generative world model that simulates the next observation in pixel space. We deliberately avoid this. 
In populated navigation, the quantities that determine whether an action is acceptable are not the visual texture of the future, but two distinct things: how nearby agents will \textit{move}, and how the navigable scene will be \textit{occupied}. We therefore decompose future prediction along these two axes --- social dynamics and scene layout --- and train a specialized predictor for each. We use the term \emph{world model} in the broad sense of an action-conditioned future predictor~\cite{bar2025navigation}, since each predictor is conditioned on a full anchor (an $L$-step sub-trajectory) and serves as a preference oracle rather than a simulator; a single anchor-conditioned prediction is sufficient for ranking candidates, and recursive frame-level rollout is unnecessary. 

Each predictor takes as input a short visual context $h_t = (I_{t-3},I_{t-2},I_{t-1},I_{t})$ and the candidate anchor $a_k$; the input is RGB so the model can read the current scene fully, but the output is restricted to a navigation-aligned target. Both predictors are pretrained, then frozen during policy refinement. 


\paragraph{State Transition World Model.} The State Transition World Model $f_\mathrm{state}$ predicts the next ego-frame relative state of each surrounding entity:
\[
\Delta \hat{p}_{t,j}(a_k), \hat{u}_{t+1,j}(a_k)=f_{\mathrm{state}}(h_t, a_k, p_{t,j}, u_{t,j}),
\]
with $\hat{p}_{t+1,j}(a_k) = p_{t,j} + \Delta \hat{p}_{t,j}(a_k)$. During pretraining, the targets are $\Delta p_{t,j}=p_{t+1,j}-p_{t,j}$ and $u_{t+1,j}$, and we minimize
\begin{equation}
\begin{split}
\mathcal{L}_{\mathrm{dyn}}
=
\mathbb{E}_{t,j}
\Big[\|\Delta \hat{p}_{t,j}-\Delta p_{t,j}\|_2^2 + \lambda_u \|\hat{u}_{t+1,j}-u_{t+1,j}\|_2^2 + \lambda_{\mathrm{dir}}
\bigl(1-\cos(\hat{u}_{t+1,j},u_{t+1,j})\bigr)
\Big].
\end{split}
\end{equation}
We add an explicit direction term because anticipating \textit{who is heading toward whom} is what makes this branch useful for ranking interaction risk. Hyperparameter values are reported in~\autoref{app:implementation}.

\paragraph{Semantic Transition World Model.}
The Semantic Transition World Model $f_\mathrm{sem}$ predicts a future segmentation map under candidate anchor
$a_k$:
\begin{equation}
\hat{S}_{t+1}(a_k) = f_{\mathrm{sem}}(h_t, a_k, \mathcal{D}_t).
\end{equation}
Two design choices distinguish this branch from a generic visual world model.

First, we predict segmentation, not RGB. The score function in Eq.~\eqref{eq:score_function} only needs to know which navigational classes the agent is about to occupy --- we use the navigation-relevant set \{sidewalk, crosswalk, road, vehicle, building, obstacle, grass\}, defined in~\autoref{app:segmentation_classes}. Predicting RGB would spend capacity on texture that the policy never reads. Predicting segmentation aligns the model's output with the quantity that actually drives semantic acceptability. 

Second, we deliberately remove pedestrians from the segmentation target. Pedestrians act as transient occluders rather than stable scene structure, and their motion is already modeled by $f_{\mathrm{state}}$. We therefore let pedestrians appear in the input image but assign their pixels an ignore label in the target, so $f_{\mathrm{sem}}$ learns to predict the underlying environmental layout --- including regions temporarily occluded by people --- rather than fleeting agent occupancy. This separation means that the two world models do not duplicate signal: $f_{\mathrm{state}}$ owns who moves where, $f_{\mathrm{sem}}$ owns what the scene affords. 

We instantiate $f_{\mathrm{sem}}$ as an EDM-style conditional denoiser~\citep{karras2022edm}. Given the clean target $S_{t+1}$, sampled noise level $\sigma>0$, Gaussian noise $\epsilon_{\mathrm{sem}}$, and noisy input $\tilde{S}_{t+1}=S_{t+1}+\sigma\epsilon_{\mathrm{sem}}$, the denoising network $D_\psi$ produces $\hat{S}^{\sigma} = D_\psi(\tilde{S}_{t+1}, \sigma, c_t(a_k))$, where $c_t(a_k)$ are the conditioning tokens built from $h_t$, $a_k$, and $\mathcal{D}_t$. We train with  
\begin{equation}
\mathcal{L}_{\mathrm{sem}}
=
\mathbb{E}_{\sigma,\epsilon_{\mathrm{sem}}}
\left[
w(\sigma)
\left\|
\hat{S}^{\sigma} - S_{t+1}
\right\|_2^2
+
\lambda_{\mathrm{LPIPS}}
\operatorname{LPIPS}
\left(
\hat{S}^{\sigma},
S_{t+1}
\right)
\right],
\end{equation}
where $w(\sigma)$ is the EDM noise-dependent weight and $\lambda_{\mathrm{LPIPS}}$ is a coefficient balancing the perceptual term against the reconstruction term.

\subsection{Consequence-Based Preference Refinement}
The two frozen world models give us, for each candidate anchor, a prediction of what would happen next. We turn these predictions into a preference signal and distill them into the policy. Crucially, this distillation is the only place in the pipeline where the world models are used; once refinement is complete, they are discarded.

\paragraph{Candidate group.} For each observation $o_t$, we form a group of ten candidates. Five are the highest-probability anchors under the current policy (exploitation of the supervised prior), and five are sampled uniformly at random from the rest of the vocabulary (exploration beyond the prior's preference). This avoids comparing only near-identical high-probability actions, which would make the refinement signal degenerate. 

\paragraph{Consequence score.} Each candidate $a_t^{(i)}$ receives a scalar score
\begin{equation}
r_t^{(i)}
=
G_t^{(i)}
- C_t^{(i)}
- P_{\mathrm{cur},t}^{(i)}
- P_{\mathrm{fut},t}^{(i)}
+ M_t^{(i)}.
\label{eq:score_function}
\end{equation}
Here $G_t^{(i)}$ measures goal progress, $C_t^{(i)}$ penalizes abrupt changes in motion, $P_{\mathrm{cur},t}^{(i)}$ measures immediate interaction risk, $P_{\mathrm{fut},t}^{(i)}$ measures predicted future interaction risk using the State Transition World Model, and $M_t^{(i)}$ measures semantic validity using the Semantic Transition World Model. All component weights are set to one and all components are scale-normalized; full formulas are in~\autoref{app:score_function}. 

\paragraph{Preference distillation.} Within each group, we update $\pi_\theta$ with GRPO, which prefers candidates whose score exceeds the group mean. This treats the dual world models as a \textit{training-time preference oracle}: rather than training the policy to imitate world-model rollouts or to plan against them online, we train it to \textit{prefer} the actions that the world models judge to lead to better consequences.

\paragraph{Training stages and deployment.} 
Training proceeds in three parts. First, the action prior $\pi_\theta$ is trained with $\mathcal{L}_{\mathrm{SFT}}$. Second, $f_{\mathrm{state}}$ and $f_{\mathrm{sem}}$ are pretrained with $\mathcal{L}_{\mathrm{dyn}}$ and $\mathcal{L}_{\mathrm{sem}}$. Third, the supervised policy checkpoint is refined with GRPO using the score in Eq.~\eqref{eq:score_function}, while both world models remain frozen.
At inference, both world models, the candidate-group construction, and the score function are discarded. Given $o_t$, the deployed policy returns a categorical distribution over $\mathcal{A}$ in a single forward pass; the selected anchor is executed directly. The look-ahead reasoning learned from the world models is amortized into $\pi_{\theta}$'s parameters, so its inference cost is independent of the world models' size, denoising depth, or rollout horizon.

\section{Experiments}



We compare against three navigation baselines spanning the major paradigms: GBM \citep{dutra2017gradient}, a synthetic-vision gradient-based steering model;  CCP~\citep{panayiotou2022ccp}, a configurable RL profile model; and CEDRL~\citep{panayiotou2025cedrl}, an example-driven RL method. All results are averaged over three runs per scenario. Implementation details are in \autoref{app:implementation}.

\begin{figure*}[t]
\centering
\includegraphics[width=\textwidth]{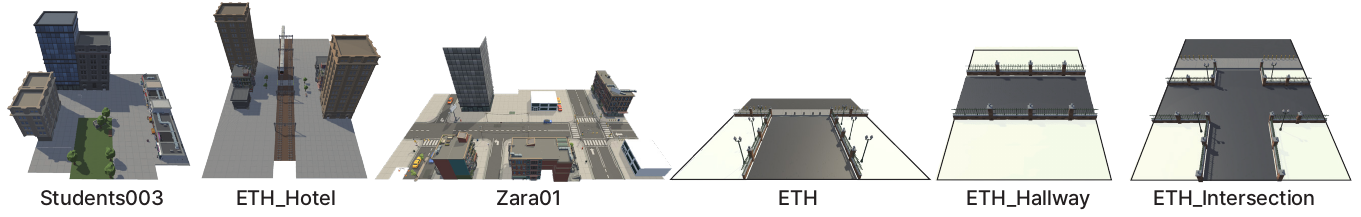}
\caption{\textbf{Overview of the training and evaluation scenes.}
We use six scenes in our experiments: Students003, ETH\_Hotel, Zara01, ETH, ETH\_Hallway, and ETH\_Intersection.
}
\label{fig:training_eval_Scene}
\end{figure*}

\begin{table*}[!t]
\centering
\small
\setlength{\tabcolsep}{6pt}
\renewcommand{\arraystretch}{1.05}
\caption{\textbf{Closed-loop results on five held-out evaluation scenarios.}
Higher is better for GSR, while lower is better for SRVR and GD.}
\label{tab:main_results}

\begin{tabular*}{\textwidth}{@{\extracolsep{\fill}}llccc@{}}
\toprule
Scene & Method & GSR $\uparrow$ & SRVR $\downarrow$ & GD $\downarrow$ \\
\midrule

\multirow{4}{*}{Zara01}
& GBM \citep{dutra2017gradient} 
& 55.6 & 21.0 & \textbf{0.6} \\
& CCP \citep{panayiotou2022ccp} 
& 33.3 & 16.0 & 2.6 \\
& CEDRL \citep{panayiotou2025cedrl} 
& 50.0 & 15.6 & 1.3 \\
& \textbf{Ours} 
& \textbf{72.2} & \textbf{7.0} & \textbf{0.6} \\
\midrule

\multirow{4}{*}{Zara01\_Group}
& GBM \citep{dutra2017gradient} 
& 38.9 & 17.5 & \textbf{1.9} \\
& CCP \citep{panayiotou2022ccp} 
& 27.8 & 19.5 & 4.8 \\
& CEDRL \citep{panayiotou2025cedrl} 
& 22.2 & 17.4 & 5.9 \\
& \textbf{Ours} 
& \textbf{77.8} & \textbf{1.2} & 2.1 \\
\midrule

\multirow{4}{*}{Students003}
& GBM \citep{dutra2017gradient} 
& 64.0 & 15.2 & 1.4 \\
& CCP \citep{panayiotou2022ccp} 
& 28.0 & 20.0 & 3.4 \\
& CEDRL \citep{panayiotou2025cedrl} 
& 37.0 & 27.1 & 1.8 \\
& \textbf{Ours} 
& \textbf{66.0} & \textbf{7.2} & \textbf{0.7} \\
\midrule

\multirow{4}{*}{ETH\_Intersection}
& GBM \citep{dutra2017gradient} 
& 72.2 & 23.5 & 2.7 \\
& CCP \citep{panayiotou2022ccp} 
& 61.1 & 13.0 & 2.3 \\
& CEDRL \citep{panayiotou2025cedrl} 
& 55.6 & 13.3 & 2.9 \\
& \textbf{Ours} 
& \textbf{83.3} & \textbf{3.9} & \textbf{0.9} \\
\midrule

\multirow{4}{*}{ETH\_Hallway}
& GBM \citep{dutra2017gradient} 
& 68.8 & 9.3 & 1.3 \\
& CCP \citep{panayiotou2022ccp} 
& 62.5 & 7.5 & 1.8 \\
& CEDRL \citep{panayiotou2025cedrl} 
& 56.3 & 19.1 & 1.5 \\
& \textbf{Ours} 
& \textbf{87.5} & \textbf{0.9} & \textbf{0.8} \\

\bottomrule
\end{tabular*}
\end{table*}

\subsection{Data and Evaluation Scenes}
We train the policy and both world models on \emph{ETH\_Hotel}, \emph{Zara01}, \emph{ETH}, and \emph{Students003} from the ETH/UCY pedestrian trajectory datasets~\citep{lerner2007crowds,pellegrini2009you}, reconstructed as Unity scenes (\autoref{fig:training_eval_Scene}). For \emph{Zara01} and \emph{Students003}, trajectories are split 8:1:1 into train, validation, and test, with only the training split used for optimization.
We evaluate on five scenarios. \emph{Zara01} and \emph{Students003} reuse training scenes with held-out trajectories. \emph{ETH\_Hallway} and \emph{ETH\_Intersection} are new layouts built from the ETH scene asset, never seen during training, to test spatial generalization, with Hallway emphasizing constrained passages and Intersection emphasizing crossing flows. \emph{Zara01\_Group} augments \emph{Zara01} with larger group configurations to test group-size generalization.

World-model inputs are short sequences of third-person RGB frames; \autoref{app:segmentation_sequences} shows example inputs with ground-truth and predicted segmentation maps.

\begin{figure*}[t]
\centering
\includegraphics[width=\textwidth]{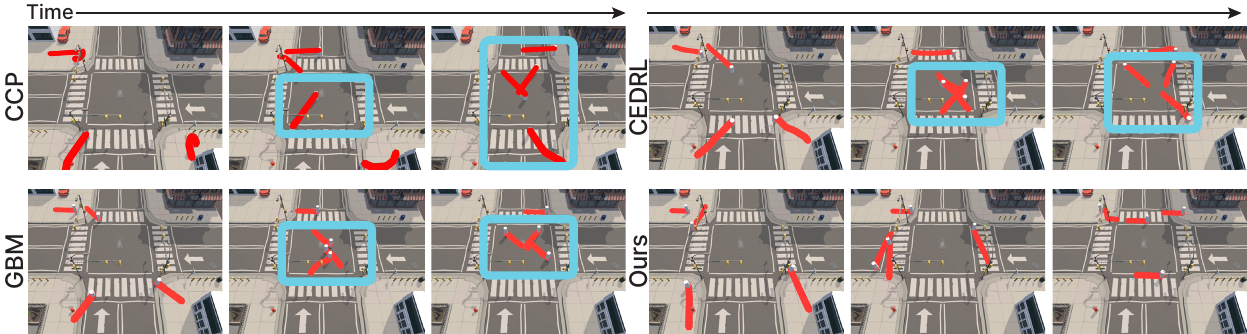}
\caption{\textbf{Crosswalk navigation comparison.}
We compare CCP, CEDRL, GBM, and our method in a road-crossing scene with a marked pedestrian crosswalk. Red curves show 
each agent's trajectory over the previous 60 frames. Blue boxes mark frames where the baselines leave the crosswalk and traverse road regions directly, indicating limited use of semantic scene cues. Our method, by contrast, remains on the crosswalk throughout the crossing, producing semantically compliant and pedestrian-like behavior. All trajectories are generated by the respective policies. In the second frame of ours, the agent moves along the crosswalk edge to avoid an oncoming pedestrian while staying on the crosswalk.}
\label{fig:crosswalk_comparison}
\end{figure*}

\subsection{Evaluation Metrics}


We evaluate along three complementary axes. Closed-loop navigation uses Goal Success Rate (GSR, percentage of agents reaching their destination collision-free), Social Region Violation Rate (SRVR,  percentage of timesteps spent in semantically inappropriate regions such as roadways), and Group Distance (GD, mean distance to the assigned group centroid), covering goal completion, scene compliance, and group coherence respectively. 
Motion fidelity uses Average and Final Displacement Error (ADE, FDE) against held-out human trajectories, measuring geometric similarity to demonstrated motion. Human perception uses 7-point Likert ratings from naive observers along motion quality, naturalness, and smoothness, which together capture perceived human-likeness. Higher is better for GSR and Likert ratings; lower is better for all other metrics.

\begin{table*}[t]
\centering
\small
\setlength{\tabcolsep}{4pt}
\renewcommand{\arraystretch}{1.05}
\caption{\textbf{Ablation study on five held-out evaluation scenarios.}
Higher is better for GSR, while lower is better for SRVR and GD.}
\label{tab:ablation_results}

\begin{tabular*}{\textwidth}{@{\extracolsep{\fill}}llccc@{}}
\toprule
Scene & Variant & GSR $\uparrow$ & SRVR $\downarrow$ & GD $\downarrow$ \\
\midrule

\multirow{7}{*}{Zara01}
& \textbf{Ours} & \textbf{72.2} & \textbf{7.0} & \textbf{0.6} \\
& w/o Dual World Models & 25.9 & 12.6 & 0.7 \\
& w/o State Transition World Model & 44.4 & 10.7 & \textbf{0.6} \\
& w/o Semantic Transition World Model & 44.4 & 9.3 & \textbf{0.6} \\
& w/o VLM & 48.1 & 10.9 & 0.9 \\
& Top-10 Candidates & 44.4 & 18.2 & 0.7 \\
& Random-10 Candidates & 22.2 & 20.2 & 0.7 \\
\midrule

\multirow{7}{*}{Zara01\_Group}
& \textbf{Ours} & \textbf{77.8} & \textbf{1.2} & 2.1 \\
& w/o Dual World Models & 16.7 & 47.0 & 1.9 \\
& w/o State Transition World Model & 27.8 & 22.3 & \textbf{1.7} \\
& w/o Semantic Transition World Model & 38.9 & 17.5 & 2.3 \\
& w/o VLM & 29.6 & 7.4 & \textbf{1.7} \\
& Top-10 Candidates & 14.8 & 42.7 & 3.2 \\
& Random-10 Candidates & 18.5 & 18.5 & 2.5 \\
\midrule

\multirow{7}{*}{Students003}
& \textbf{Ours} & \textbf{66.0} & \textbf{7.2} & \textbf{0.7} \\
& w/o Dual World Models & 44.7 & 14.4 & \textbf{0.7} \\
& w/o State Transition World Model & 47.0 & 9.5 & 0.8 \\
& w/o Semantic Transition World Model & 63.3 & 8.9 & 0.8 \\
& w/o VLM & 60.0 & 10.6 & 0.8 \\
& Top-10 Candidates & 61.3 & 8.3 & 1.0 \\
& Random-10 Candidates & 57.3 & 8.5 & 0.8 \\
\midrule

\multirow{7}{*}{ETH\_Intersection}
& \textbf{Ours} & \textbf{83.3} & \textbf{3.9} & 0.9 \\
& w/o Dual World Models & 40.7 & 13.5 & 1.1 \\
& w/o State Transition World Model & 51.9 & 10.6 & 1.0 \\
& w/o Semantic Transition World Model & 66.7 & 6.2 & 0.9 \\
& w/o VLM & 63.0 & 7.4 & 1.1 \\
& Top-10 Candidates & 55.6 & 13.9 & 1.0 \\
& Random-10 Candidates & 44.4 & 14.2 & \textbf{0.8} \\
\midrule

\multirow{7}{*}{ETH\_Hallway}
& \textbf{Ours} & \textbf{87.5} & \textbf{0.9} & 0.8 \\
& w/o Dual World Models & 58.3 & 9.9 & 1.0 \\
& w/o State Transition World Model & 58.3 & 8.7 & 0.8 \\
& w/o Semantic Transition World Model & 70.8 & 13.7 & \textbf{0.6} \\
& w/o VLM & 66.7 & 16.9 & 0.9 \\
& Top-10 Candidates & 62.5 & 3.4 & 0.7 \\
& Random-10 Candidates & 70.8 & 12.5 & 1.2 \\

\bottomrule
\end{tabular*}
\end{table*}

\begin{table*}[t]
\centering
\small
\setlength{\tabcolsep}{4pt}
\renewcommand{\arraystretch}{1.05}
\caption{\textbf{Comparison with simpler pipelines and an online world-model planner.}
Trajectory predictors are combined with an MPC planner~\citep{poddar2023crowd}. Results are averaged over the five held-out evaluation scenarios; per-scene results are in \autoref{app:extra_baselines}. Higher is better for GSR, while lower is better for SRVR and GD. OnlineWM runs both world models at every decision; bold marks the best result among the remaining methods.}
\label{tab:extra_baselines}
\begin{tabular*}{\textwidth}{@{\extracolsep{\fill}}lccc@{}}
\toprule
Method & GSR $\uparrow$ & SRVR $\downarrow$ & GD $\downarrow$ \\
\midrule
OnlineWM & 80.6 & 3.2 & 0.9 \\
\midrule
Constant Velocity~\citep{scholler2020constant} & 56.1 & 15.6 & 2.2 \\
Social-LSTM~\citep{alahi2016sociallstm} & 51.9 & 15.0 & 2.1 \\
Social-GAN~\citep{gupta2018socialgan} & 53.2 & 15.0 & 2.2 \\
Trajectron++~\citep{salzmann2020trajectron} & 54.4 & 17.0 & 2.1 \\
Gen-Safe-Nav~\citep{yao2025towards} & 42.9 & 14.1 & 2.3 \\
\textbf{Ours} & \textbf{77.4} & \textbf{4.0} & \textbf{1.0} \\
\bottomrule
\end{tabular*}
\end{table*}

\subsection{Experimental Results and Discussion}
We organize the results around our three contributions: (i) consequence-aware action selection improves closed-loop behavior over reactive and learning-based baselines, (ii) the two world models capture complementary failure modes, and (iii) the resulting policy is more human-like in both trajectory geometry and human perception, while requiring only a single forward pass at inference. The two world models that drive refinement are themselves accurate on held-out data ---  87.0\% mIoU and 99.3\% pixel accuracy for the semantic branch, 0.18\,m position ADE and 20.8$^{\circ}$ heading error for the state branch (\autoref{app:world_model_performance}) --- so the preference scores guiding GRPO rest on reliable predictions. 

\paragraph{Closed-loop navigation across five held-out scenarios.}
\autoref{tab:main_results} reports closed-loop performance on the five evaluation scenes, and \autoref{fig:crosswalk_comparison} shows a qualitative example.
Across five evaluation scenes, our method achieves the best GSR and lowest SRVR in every scene, and the lowest or tied-lowest GD in four out of five scenes. Magnitudes matter: on Zara01\_Group, SRVR drops from 17.4 (next-best) to 1.2, an order-of-magnitude reduction, with GSR doubling (38.9 $\rightarrow$ 77.8); ETH\_Hallway shows a similar pattern (SRVR 7.5$\rightarrow$0.9, GSR 68.8$\rightarrow$87.5), as does the multi-flow ETH\_Intersection (SRVR 13.0$\rightarrow$3.9, GSR 72.2$\rightarrow$83.3). Crucially, lower violations do not come from conservatism: GSR is higher than the strongest baseline on every scene.
\autoref{fig:crosswalk_comparison} illustrates this qualitatively: at a marked crosswalk, the baselines leave the crosswalk and cut across the road, whereas our agent stays on it throughout the crossing.
This supports our central claim that ranking locally plausible candidates by their predicted consequences yields navigation that is simultaneously safer, more compliant, and more goal-directed than baselines that act on the current state alone.


\paragraph{Complementary roles of the two world models.}
\autoref{tab:ablation_results} isolates the contribution of each design choice. Removing both world models causes the largest average degradation (mean GSR 77.4 $\rightarrow$ 37.3, mean SRVR 4.0 $\rightarrow$ 19.5) and worsens both metrics on every scene, most starkly on Zara01\_Group, where SRVR rises from 1.2 to 47.0 and GSR collapses from 77.8 to 16.7. This confirms that the supervised action prior alone cannot distinguish locally plausible anchors with different downstream consequences. Additional qualitative examples are provided in \autoref{app:additional_qualitative}.

The two world models cover different kinds of mistakes, as the single-branch ablations make clear. Both branches are themselves accurate on held-out data (87.0\% semantic mIoU, 0.18 m state ADE; full diagnostics in~\autoref{app:world_model_performance}), so the differences below reflect \emph{which signal} matters where, not differences in prediction quality. 
The State Transition World Model is most critical when interaction risk dominates: in ETH\_Intersection, removing it drops GSR from 83.3 to 51.9 (a larger degradation than removing the semantic branch, which gives 66.7), reflecting that crossing flows are governed primarily by \textit{who is heading toward whom}. Conversely, the Semantic Transition World Model is most critical when the navigable layout is the binding constraint: in ETH\_Hallway, removing it raises SRVR from 0.9 to 13.7 (versus 8.7 when the state branch is removed), reflecting that constrained passages punish semantically invalid footprints more than mis-predicted neighbor trajectories. Removing either branch lowers GSR and raises SRVR in every scene, so the two branches are complementary; the state branch has the larger effect in most scenes.

The candidate-construction ablation (Top-10 vs. Random-10 vs. our top-5 + random-5) shows that the GRPO refinement also relies on a sufficiently diverse comparison set: pure exploitation collapses on Zara01\_Group (GSR 14.8) while pure exploration is unstable on ETH\_Intersection (GSR 44.4). The mixed strategy works because the two extremes fail in opposite ways: top-only gives candidates too similar to learn from, random-only gives candidates too irrelevant to matter. Mixing exposes the policy to alternatives that are nearby enough to be reachable yet different enough to be informative.
The w/o VLM result shows that replacing the pretrained VLM with a task-trained Transformer/ViT consistently lowers GSR and raises SRVR, indicating that pretrained multimodal grounding provides useful scene-level cues beyond visual conditioning alone.

\paragraph{Is a world model necessary?}
To test whether a simpler pipeline suffices, we pair trajectory predictors~\citep{scholler2020constant,alahi2016sociallstm,gupta2018socialgan,salzmann2020trajectron} with an MPC planner~\citep{poddar2023crowd}, and compare against Gen-Safe-Nav~\citep{yao2025towards}, which uses conformal prediction to account for forecast uncertainty. As \autoref{tab:extra_baselines} shows, all of these alternatives reach clearly lower goal success and higher violation rates than our method, suggesting that more accurate forecasting alone does not lead to safer navigation, consistent with~\cite{poddar2023crowd}. We also evaluate an online planner that runs both world models at every decision (OnlineWM). Our distilled policy recovers most of its performance while requiring only a single forward pass (65\,ms per decision, compared with 1.2\,s for OnlineWM). Input ablations (\autoref{app:input_ablation}) further show that visual and state inputs are complementary, and results on the standard ETH/UCY forecasting benchmark (\autoref{app:standard_benchmark}) show that the policy remains competitive outside our Unity reconstructions.

\paragraph{Human-likeness and inference cost.}
Three lines of evidence indicate that the refined policy produces human-like motion while retaining a single-pass inference profile. First, on motion-fidelity metrics that compare generated rollouts against held-out human trajectories (\autoref{app:motion_fidelity}), our method achieves the lowest ADE on both evaluated scenes (0.2064 m on Zara01, 0.1994 m on Students003) and the lowest FDE on Zara01 (0.3897 m), with Students003 FDE close to that of the best baseline. Second, in a user study with 20 naive participants rating anonymized rollouts on 7-point Likert scales (\autoref{app:user_study}), our method receives the highest average rating across all three perceptual axes --- motion quality, naturalness, and smoothness --- over GBM, CCP, and CEDRL. Third, the look-ahead reasoning is amortized at training time only: per-candidate scoring during refinement takes 0.115\,s for the semantic branch and 1.9\,ms for the state branch (\autoref{app:runtime_compute_memory}), but both world models are removed at deployment, leaving only the policy's single anchor-classification step at 65\,ms per decision. 
Together, the geometric, perceptual, and computational evidence indicates that consequence-aware preference distillation produces motion that is human-aligned and naive-observer-preferred, without paying any look-ahead cost at inference.

\section{Limitations}
\label{sec:limitations}
Our policy selects from a fixed vocabulary of $K=100$ short-horizon anchors; motions outside this vocabulary can only be approximated, and longer horizons would require larger or learned vocabularies. Both world models make a single anchor-conditioned prediction, so they cannot capture long-range effects such as delayed yielding or group reformation over several seconds. When the two world models disagree, the policy follows their unit-weighted sum in Eq.~\eqref{eq:score_function} rather than an explicit arbitration rule; a learned or context-dependent weighting may resolve such conflicts better. All closed-loop results come from Unity reconstructions of ETH/UCY scenes, and we have not validated the policy on a real robot under sensor noise, localization error, or actuation delay. Finally, the pipeline is more complex and more costly to train than prediction-plus-planning alternatives; when scenes carry few semantic constraints or training compute is limited, a simpler predictor with MPC may be the more practical choice. Further failure cases are discussed in \autoref{app:failure_cases_limitations}.

\section{Conclusion}
The shape of ``what to do next'' in a populated scene is governed by what each action will cause, not by what it currently avoids, and the standard remedy of running a world model online at every step is a cost most deployments cannot afford. We close this gap with three intertwined moves: treating world models as training-time preference oracles rather than online planners, decomposing future prediction along the two axes that actually drive navigation, and aligning each branch with the failure mode it is best placed to catch. Together, these yield a policy that is simultaneously safer, more socially and group-coherent, more goal-directed, and perceived as more human-like, while paying no foresight cost at inference. 

We see this as one instance of a broader direction. As embodied agents are built on large pretrained backbones, the question is no longer whether models can simulate the future, but how foresight can be made cheap and targeted enough to act on. Amortizing rollouts into the policy is one move; restricting prediction to the decision-relevant slice of the future is another. Natural next steps include learning the motion vocabulary jointly with the policy and extending to scenarios involving deferred coordination, where hierarchical world models may help. More broadly, the same principle should extend to settings where actions are rich and consequences are not visible from the current observation, and where the world models themselves keep improving without the deployed agent ever having to wait for them.

\newpage

\begin{ack}
This work was partly supported by Institute of Information \& Communications Technology Planning \& Evaluation (IITP) under the Leading Generative AI Human Resources Development (IITP-2026-RS-2025-25441313, Contribution: 10\%) grant funded by the Korea government (MSIT);
by ICT Creative Consilience Program through the Institute of Information \& Communications Technology Planning \& Evaluation (IITP) grant funded by the Korea government (MSIT) (IITP-2026-RS-2020-II201819, Contribution: 45\%);
by the Institute of Information \& Communications Technology Planning \& Evaluation (IITP) grant funded by the Korea government (MSIT) (IITP-2026-RS-2026-25615490, AI Star Fellowship Support Program, Contribution: 45\%).
\end{ack}







\bibliographystyle{plainnat}
\bibliography{neurips_2026}







\clearpage
\appendix

\section{Implementation Details}
\label{app:implementation}

We use Qwen3-VL-2B-Instruct~\citep{bai2025qwen3} as the policy backbone, frozen except for LoRA adapters (rank 64, $\alpha=16$, dropout 0.05) --- chosen so that the visual prior does not drift during anchor classification.
We train the anchor classifier with cross-entropy for 10 epochs at learning rate $3\times10^{-5}$, batch size 32 (accumulation 4), cosine schedule with 5\% warmup, weight decay 0.02, and gradient checkpointing.
GRPO refines the supervised checkpoint for 10 epochs at learning rate $2\times10^{-5}$ with both world models frozen. Each candidate group contains 10 anchors: the top five under the current policy and five sampled uniformly from the remainder. The semantic transition model runs 25 denoising steps per candidate. World-model context uses three history frames; the state branch uses $\lambda_u=1.0$, $\lambda_{\mathrm{dir}}=2.0$, and the semantic branch uses $\lambda_{\mathrm{LPIPS}}=0.1$. All five score components in Eq.~\eqref{eq:score_function} are summed with unit weights, deliberately avoiding scenario-specific reward tuning. For training, we used eight NVIDIA A100 GPUs, while all inference experiments were conducted on a single NVIDIA RTX 6000 Ada GPU.

\section{Score Function Details}
\label{app:score_function}

This appendix gives the detailed score terms used in Eq.~\eqref{eq:score_function}. For candidate anchor $a_t^{(i)}=\{y_{i,0},\ldots,y_{i,L}\}$, $y_{i,L}$ is the final displacement point, $b_i=y_{i,1}$ is the first executed displacement, and $\bar{v}_i=y_{i,1}-y_{i,0}$ is the first-step velocity implied by the anchor. The operator $\operatorname{clip}(x,l,u)$ clamps scalar $x$ into $[l,u]$.

\paragraph{Goal progress and smoothness.}
The goal-progress score is the normalized reduction in distance to the ego-relative goal $g_t$:
\begin{equation}
G_t^{(i)}
=
\operatorname{clip}
\left(
\frac{\|g_t\|_2-\|g_t-y_{i,L}\|_2}{\|g_t\|_2+\epsilon},
-1,1
\right).
\end{equation}
The smoothness cost penalizes abrupt changes from the previous ego velocity $v_{t-1}$:
\begin{equation}
C_t^{(i)}
=
\operatorname{clip}
\left(
\frac{\|\bar{v}_i-v_{t-1}\|_2^2}
{2(\|\bar{v}_i\|_2^2+\|v_{t-1}\|_2^2)+\epsilon},
0,1
\right),
\end{equation}
with $C_t^{(i)}=0$ when both velocity magnitudes are numerically zero.

\paragraph{Immediate and predicted future states.}
For a surrounding entity with current relative position $p_{t,j}$ and relative velocity $u_{t,j}$, the immediate state after applying candidate $a_t^{(i)}$ is approximated as
\begin{equation}
\tilde{p}_{t,j}^{(i)} = p_{t,j} + (u_{t,j}+v_t) - b_i,
\qquad
\tilde{u}_{t,j}^{(i)} = (u_{t,j}+v_t) - \bar{v}_i.
\end{equation}
The predicted future state is obtained from the State Transition World Model:
\begin{equation}
p_{t+1,j}^{(i)} = p_{t,j} + \Delta\hat{p}_{t,j}(a_t^{(i)}),
\qquad
u_{t+1,j}^{(i)} = \hat{u}_{t+1,j}(a_t^{(i)}).
\end{equation}

\paragraph{Interaction costs.}
The interaction score uses three costs: crowd pressure, proxemic cost, and collision risk. For pedestrian positions $P=\{p_j\}_{j=1}^{n}$ and pedestrian velocities $U=\{u_j\}_{j=1}^{n}$, crowd pressure is
\begin{equation}
\mathcal{I}(P,U)
=
\operatorname{clip}
\left(
\frac{\sum_{j=1}^{n}\exp\left(-\|p_j\|_2^2/(2\sigma_\rho^2)\right)}
{N_{\max}},
0,1
\right)
\operatorname{clip}
\left(
\frac{
\frac{1}{n}\sum_{j=1}^{n}\|u_j-\bar{u}\|_2^2
}{
\frac{1}{n}\sum_{j=1}^{n}\|u_j\|_2^2+\epsilon
},
0,1
\right),
\end{equation}
where $\sigma_\rho=2.5$, $N_{\max}=15$, and $\bar{u}=n^{-1}\sum_{j=1}^{n}u_j$. If no pedestrian is present, the pressure term is set to zero.

The proxemic cost uses the same pedestrian positions and group indicators $q_j\in\{0,1\}$:
\begin{equation}
\mathcal{C}(P,q)
=
\operatorname{clip}
\left(
\frac{1}{N_{\max}}
\sum_{j=1}^{n}
w_j \exp\left(-\|p_j\|_2^2/(2\sigma_p^2)\right),
0,1
\right),
\qquad
w_j=
\begin{cases}
0.5, & q_j=1,\\
1, & q_j=0,
\end{cases}
\end{equation}
where $\sigma_p=1.25$. The smaller weight for $q_j=1$ reduces the penalty for pedestrians in the same group as the controlled agent.

Collision risk is computed for both pedestrians and dynamic objects. For positions $P=\{p_j\}_{j=1}^{n}$ and velocities $U=\{u_j\}_{j=1}^{n}$, the time-to-collision proxy is
\begin{equation}
T_j = -\frac{p_j^\top u_j}{\|u_j\|_2^2+\epsilon}.
\end{equation}
The raw collision score with entity-type scale $s$ is
\begin{equation}
\mathcal{R}_{\mathrm{raw}}(P,U;s)
=
\sum_{j:T_j>0}
s\exp(-T_j/\tau),
\end{equation}
where $\tau=2.0$. Pedestrians use $s=1$, and dynamic objects use $s=s_{\mathrm{obj}}=1.2$. The normalized collision risk is
\begin{equation}
\mathcal{R}(P^{\mathrm{ped}},U^{\mathrm{ped}},P^{\mathrm{obj}},U^{\mathrm{obj}})
=
\operatorname{clip}
\left(
\frac{
\mathcal{R}_{\mathrm{raw}}(P^{\mathrm{ped}},U^{\mathrm{ped}};1)
+
\mathcal{R}_{\mathrm{raw}}(P^{\mathrm{obj}},U^{\mathrm{obj}};s_{\mathrm{obj}})
}{
N_{\max}+s_{\mathrm{obj}}M_{\max}+\epsilon
},
0,1
\right),
\end{equation}
where $M_{\max}=10$ is the maximum number of dynamic objects.

The immediate interaction penalty is
\begin{equation}
P_{\mathrm{cur},t}^{(i)}
=
I_{\mathrm{cur},t}^{(i)}
+ C_{\mathrm{cur},t}^{(i)}
+ R_{\mathrm{cur},t}^{(i)},
\end{equation}
where $I_{\mathrm{cur},t}^{(i)}$, $C_{\mathrm{cur},t}^{(i)}$, and $R_{\mathrm{cur},t}^{(i)}$ evaluate $\mathcal{I}$, $\mathcal{C}$, and $\mathcal{R}$ on the immediate states $\tilde{p}_{t,j}^{(i)}$ and $\tilde{u}_{t,j}^{(i)}$. The future interaction penalty is
\begin{equation}
P_{\mathrm{fut},t}^{(i)}
=
I_{\mathrm{fut},t}^{(i)}
+ C_{\mathrm{fut},t}^{(i)}
+ R_{\mathrm{fut},t}^{(i)},
\end{equation}
where $I_{\mathrm{fut},t}^{(i)}$, $C_{\mathrm{fut},t}^{(i)}$, and $R_{\mathrm{fut},t}^{(i)}$ evaluate the same costs on the predicted future states $p_{t+1,j}^{(i)}$ and $u_{t+1,j}^{(i)}$.

\paragraph{Semantic validity.}
The semantic score is computed from the predicted segmentation $\hat{S}_{t+1}(a_t^{(i)})$. A triangular local footprint is placed in the image, and $\rho_c^{(i)}$ denotes the percentage of footprint pixels assigned to semantic class $c$. Define $\phi_{\geq}(x,\tau,a,b)$ to return $a$ if $x\geq\tau$ and $b$ otherwise. The raw semantic score is
\begin{equation}
\begin{split}
M_{\mathrm{raw},t}^{(i)}
=&
\phi_{\geq}(\rho_{\mathrm{sidewalk}}^{(i)}+\rho_{\mathrm{crosswalk}}^{(i)},70,1,-1)
+ \phi_{\geq}(\rho_{\mathrm{road}}^{(i)},70,-2,1) \\
&+ \phi_{\geq}(\rho_{\mathrm{grass}}^{(i)},25,-1,1)
+ \phi_{\geq}(\rho_{\mathrm{vehicle}}^{(i)},5,-10,1) \\
&+ \phi_{\geq}(\rho_{\mathrm{building}}^{(i)},5,-10,1)
+ \phi_{\geq}(\rho_{\mathrm{obstacle}}^{(i)},5,-10,1).
\end{split}
\end{equation}
The normalized semantic score is
\begin{equation}
M_t^{(i)}
=
\operatorname{clip}
\left(
\frac{M_{\mathrm{raw},t}^{(i)}}{34},
-1,1
\right),
\end{equation}
where $34$ is the maximum absolute rule score.

The reward and penalty values in the semantic score were designed according to the relative safety criticality of each semantic class. In particular, vehicle, building, and obstacle regions represent physical objects or structures for which contact must be strictly avoided. Since these classes may occupy a relatively small portion of the local footprint compared with terrain-related classes such as grass, road, sidewalk, and crosswalk, even a small overlap can indicate an unsafe or infeasible motion. Therefore, they are assigned the largest penalties. Road regions are also treated as invalid for pedestrian navigation, but they do not necessarily imply direct physical collision with a static or dynamic object; thus, they are assigned an intermediate penalty. Grass, sidewalk, and crosswalk are given lower penalty or reward magnitudes because they mainly encode terrain preference and pedestrian-validity rather than hard collision constraints. Accordingly, the penalty severity follows the hierarchy
\[
\mathrm{vehicle}, \mathrm{building}, \mathrm{obstacle}
>
\mathrm{road}
>
\mathrm{grass}, \mathrm{sidewalk}, \mathrm{crosswalk}.
\]
The remaining scalar values, including class-specific thresholds and reward/penalty magnitudes, were determined empirically through hyperparameter tuning while preserving this relative ordering of semantic penalty severity.

\section{Segmentation Classes}
\label{app:segmentation_classes}

The Semantic Transition World Model predicts a compact navigation-oriented segmentation map rather than a photorealistic RGB frame. 
Each semantic region is represented by a discrete tag and a fixed Hex color code. 
The Hex code is used only as a deterministic label encoding and for visualization; during training and scoring, segmentation maps are converted into semantic class indices or one-hot masks.

\paragraph{Class vocabulary.}
We map all semantic annotations into the class set shown in \autoref{tab:semantic_classes}. 
The vocabulary covers pedestrian-walkable areas, vehicle-oriented regions, and static structures. 
Unlike generic scene-parsing taxonomies, this vocabulary is chosen to match navigation-relevant distinctions: whether a candidate motion remains on pedestrian-compatible space, approaches a dynamic or static obstacle, or enters a road-like danger zone.

\begin{table*}[t]
\centering
\caption{\textbf{Segmentation classes used by the Semantic Transition World Model.} Each class is identified by a semantic tag and a fixed Hex color code. The score role describes how the class is treated by the semantic-validity score in \autoref{app:score_function}.}
\label{tab:semantic_classes}
\small
\begin{tabular}{p{0.14\linewidth}|p{0.22\linewidth}|p{0.11\linewidth}|p{0.39\linewidth}}
\toprule
Semantic tag & Navigation meaning & Hex & Role in semantic score \\
\midrule
\texttt{sidewalk} 
& Pedestrian sidewalk or primary outdoor walking surface. 
& \texttt{\#804080} 
& Positive. A high sidewalk ratio indicates that the candidate anchor remains on a pedestrian-compatible walking region. \\
\midrule
\texttt{crosswalk} 
& Marked pedestrian crossing region. 
& \texttt{\#FFFFFF} 
& Positive. Treated as a valid walking region and grouped with sidewalk for walkability scoring. \\
\midrule
\texttt{road} 
& Vehicle-oriented road surface. 
& \texttt{\#808080} 
& Negative. A large road ratio indicates unsafe or semantically inappropriate pedestrian motion unless the region is explicitly labeled as crosswalk. \\
\midrule
\texttt{obstacle} 
& Static obstacle such as a pole, barrier, fence, furniture, or other non-traversable object. 
& \texttt{\#FF8C00} 
& Strongly negative. Footprint overlap indicates physically invalid or unsafe navigation. \\
\midrule
\texttt{building} 
& Building, wall, or other non-traversable static structure. 
& \texttt{\#464646} 
& Strongly negative. Overlap indicates an invalid trajectory through static structure. \\
\midrule
\texttt{vehicle} 
& Moving or parked vehicle. 
& \texttt{\#00008E} 
& Strongly negative. Even a small vehicle overlap indicates high collision or scene-validity risk. \\
\midrule
\texttt{grass} 
& Grass or off-path terrain. 
& \texttt{\#6B8E23} 
& Mildly negative. Penalized when it occupies a substantial fraction of the local footprint because it is not the primary pedestrian walking surface. \\
\midrule
\texttt{unknown} 
& Undefined object, unlabeled pixel, or background. 
& \texttt{\#000000} 
& Neutral. Ignored in semantic-validity scoring unless explicitly replaced by a known non-walkable class. \\
\bottomrule
\end{tabular}
\end{table*}

\begin{figure*}[t]
\centering
\includegraphics[width=\textwidth]{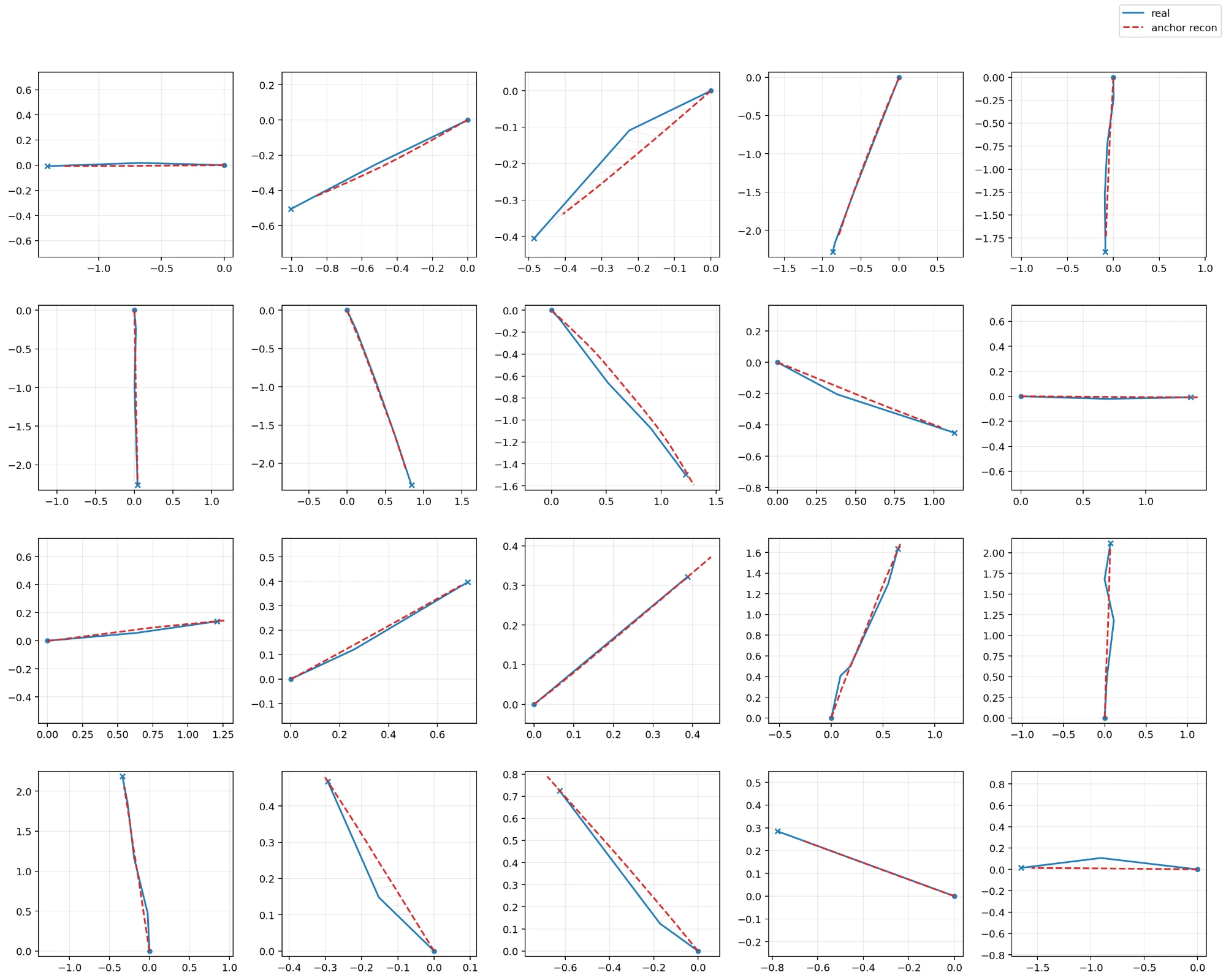}
\caption{\textbf{Real trajectory clips and nearest-anchor reconstructions.}
We randomly sample 20 trajectory clips and visualize each real trajectory against its nearest anchor reconstruction from the $K=100$ vocabulary. 
Blue solid lines denote real trajectory clips, and red dashed lines denote anchor reconstructions. 
The examples show that the 100-anchor vocabulary preserves the main direction, scale, and endpoint tendency of short-horizon pedestrian motion, supporting its use as a compact discrete action space.}
\label{fig:anchor_recon_samples}
\end{figure*}

\section{Anchor Vocabulary Reconstruction Analysis}
\label{app:anchor_vocabulary}

The purpose of the anchor vocabulary is not to reproduce every continuous trajectory exactly, but to provide a compact discrete action space that still preserves the main short-horizon motion patterns observed in real pedestrian data. 
In all experiments, we use $K=100$ anchors, where each anchor represents a 20-frame short-horizon motion primitive extracted from the ETH/UCY trajectory data.

To examine whether this vocabulary is sufficiently expressive, we randomly sample 20 trajectory clips from the dataset and compare each real trajectory with its nearest anchor reconstruction. 
\autoref{fig:anchor_recon_samples} shows the comparison. 
The blue solid curves indicate the original real trajectory clips, and the red dashed curves indicate the corresponding anchor reconstructions. 
Across the sampled examples, the reconstructed anchors closely follow the direction, scale, and endpoint tendency of the real trajectories. 
Although small deviations appear in some clips, especially for rare or abrupt motions, the nearest anchors generally preserve the dominant movement pattern needed for short-horizon action selection.

This result supports the use of $K=100$ as a practical vocabulary size. 
A smaller vocabulary would make the action space more compact, but it would merge distinct behaviors such as straight walking, lateral movement, turning, and speed variation into overly coarse motion primitives. 
On the other hand, a much larger vocabulary would increase the number of action classes, make supervised anchor classification more sparse, and increase the cost of candidate comparison during world-model-guided refinement. 
The reconstruction examples in \autoref{fig:anchor_recon_samples} indicate that $K=100$ provides a good balance: it is expressive enough to approximate diverse short-horizon pedestrian motions, while remaining compact enough for efficient policy learning and candidate scoring.

Importantly, the anchor representation is used as an action abstraction rather than as a final trajectory predictor. 
During policy learning, the model only needs to select a plausible short-horizon motion primitive from the current observation. 
During refinement, the dual world models compare candidate anchors according to their predicted social and semantic consequences. 
Therefore, exact point-wise reconstruction is less important than preserving behaviorally meaningful motion modes. 
The close agreement between real clips and anchor reconstructions in the random samples suggests that the 100-anchor vocabulary satisfies this requirement.

\section{World Model Performance and Diagnostics}
\label{app:world_model_performance}

This appendix reports diagnostic test performance for the two frozen consequence predictors used during policy refinement: the State Transition World Model and the Semantic Transition World Model. 
The goal of this evaluation is not to benchmark a general-purpose simulator, but to verify that each branch is accurate on the decision-relevant target it contributes to candidate-anchor scoring.

\paragraph{Diagnostic rationale for the dual design.}
We evaluate the two world models separately because they answer different questions during preference construction. 
The State Transition World Model asks \emph{who will move where} after a candidate anchor is executed. 
It therefore predicts future relative positions, velocities, speeds, and headings of nearby pedestrians and dynamic objects, which are used to estimate interaction risk. 
The Semantic Transition World Model asks \emph{what navigational affordance the agent's future footprint will occupy}. 
It therefore predicts a navigation-oriented semantic layout, which is used to evaluate whether a candidate anchor remains on pedestrian-compatible regions or moves into invalid regions such as roads, vehicles, buildings, or obstacles.

This separation is especially important in populated scenes. 
Pedestrians are dynamic agents and transient occluders: they may temporarily cover a sidewalk or crosswalk, but they do not change the underlying walkability of that region. 
If pedestrians were included as ordinary segmentation targets, the semantic branch would be forced to solve two different problems at once: predicting persistent scene layout and forecasting every future pedestrian mask. 
This would duplicate the role of the state branch and introduce noisy supervision, because small errors in future pedestrian placement can dominate segmentation loss without improving semantic footprint scoring. 
We therefore let pedestrians appear in the RGB input, but remove them from the segmentation target by assigning their pixels to an ignore label. 
The semantic branch learns the underlying environmental layout, including walkable regions temporarily occluded by people, while the state branch models dynamic pedestrian motion.

The following diagnostics are therefore branch-specific. 
For the semantic branch, we report class-aware segmentation metrics that measure layout fidelity. 
For the state branch, we report position, velocity, speed, and heading errors that measure short-horizon interaction prediction quality.

\paragraph{Semantic Transition World Model.}
The Semantic Transition World Model predicts the next navigation-oriented segmentation map. 
Since this model is used to evaluate whether a candidate anchor moves the agent toward semantically valid or invalid regions, we report class-aware segmentation metrics rather than RGB reconstruction or perceptual losses. 
The reported metrics include pixel accuracy, mean IoU, Dice/F1, and frequency-weighted IoU. 
Pixel accuracy measures overall pixel-wise correctness, mIoU evaluates class-balanced overlap quality, Dice/F1 measures region overlap, and frequency-weighted IoU summarizes IoU while accounting for class frequency.

\begin{table*}[t]
\centering
\caption{\textbf{Semantic prediction performance.}
Class-aware test metrics for the Semantic Transition World Model.}
\label{tab:semantic_wm_test_metrics}
\footnotesize
\setlength{\tabcolsep}{0pt}
\begin{tabular*}{\textwidth}{@{\extracolsep{\fill}}lcccc@{}}
\toprule
Method
& Pixel Acc. $\uparrow$
& mIoU $\uparrow$
& Dice/F1 $\uparrow$
& FWIoU $\uparrow$ \\
\midrule
EDM
& 99.2848\%
& 87.0020\%
& 89.0814\%
& 98.6099\% \\
DDPM
& 97.7863\% 
& 86.5421\% 
& 88.2132\% 
& 95.4289\%  \\
\bottomrule
\end{tabular*}
\end{table*}

As shown in Table~\ref{tab:semantic_wm_test_metrics}, the EDM-style semantic transition model achieves the best overall performance, with 99.2848\% pixel accuracy, 87.0020\% mIoU, 89.0814\% Dice/F1, and 98.6099\% frequency-weighted IoU. 
Compared with the DDPM variant, EDM improves pixel accuracy and frequency-weighted IoU while also slightly improving mIoU and Dice/F1. 
These results indicate that the semantic branch preserves both dominant layout structure and class-level region overlap, supporting its use as a footprint-based semantic consequence estimator.

\paragraph{State Transition World Model.}
The State Transition World Model predicts the next relative position and velocity of nearby pedestrians and dynamic objects. 
This prediction is used to estimate future interaction risk, including possible collisions, crowd pressure, and proxemic discomfort. 
We therefore evaluate the model with position, velocity, speed, and heading errors that directly reflect short-horizon state prediction quality.

\begin{table*}[t]
\centering
\caption{\textbf{State prediction performance.}
Test metrics for future relative position and velocity prediction.}
\label{tab:state_wm_test_metrics}
\footnotesize
\setlength{\tabcolsep}{0pt}
\begin{tabular*}{\textwidth}{@{\extracolsep{\fill}}cccccc@{}}
\toprule
Pos. ADE $\downarrow$
& Pos. RMSE $\downarrow$
& Vel. MAE $\downarrow$
& Vel. RMSE $\downarrow$
& Speed MAE $\downarrow$
& Head. Err. $\downarrow$ \\
\midrule
0.1846
& 0.2481
& 0.3354
& 0.4414
& 0.2096
& 20.77 \\
\bottomrule
\end{tabular*}
\end{table*}

As shown in Table~\ref{tab:state_wm_test_metrics}, the State Transition World Model achieves a position ADE of 0.1846\,m and a position RMSE of 0.2481\,m, showing accurate short-horizon relative position prediction. 
For motion prediction, it obtains a velocity MAE of 0.3354\,m/s, a velocity RMSE of 0.4414\,m/s, and a speed MAE of 0.2096\,m/s. 
The heading error is 20.77 degrees, indicating that the model also captures the directional tendency of nearby moving entities. 
These results show that the state branch provides sufficiently accurate future relative-state estimates for interaction-aware candidate scoring, complementing the semantic branch rather than duplicating it.

Together, these diagnostics support the dual-world-model decomposition. 
The semantic branch provides reliable predictions of persistent navigation affordances, including walkable regions that may be temporarily occluded by pedestrians, while the state branch provides reliable short-horizon estimates of nearby dynamic entities. 
This division gives the preference score two non-overlapping sources of foresight: semantic validity from predicted layout and interaction risk from predicted motion. 
The main ablation results further test the behavioral necessity of using both signals during policy refinement.

\begin{figure*}[t]
\centering
\includegraphics[width=\textwidth]{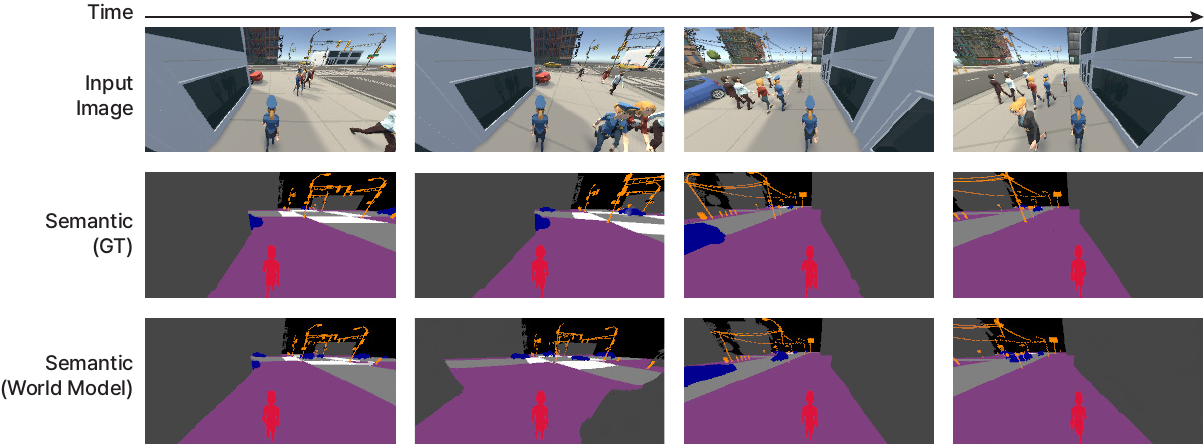}
\caption{\textbf{Segmentation image sequence from Zara01.} Columns are ordered in time. Rows show the third-person RGB input, the ground-truth segmentation, and the segmentation predicted by the Semantic Transition World Model. The red silhouette marks the controlled agent; other pedestrians are excluded from the segmentation target.}
\label{fig:SemanticImages_Zara}
\end{figure*}

\begin{figure*}[t]
\centering
\includegraphics[width=\textwidth]{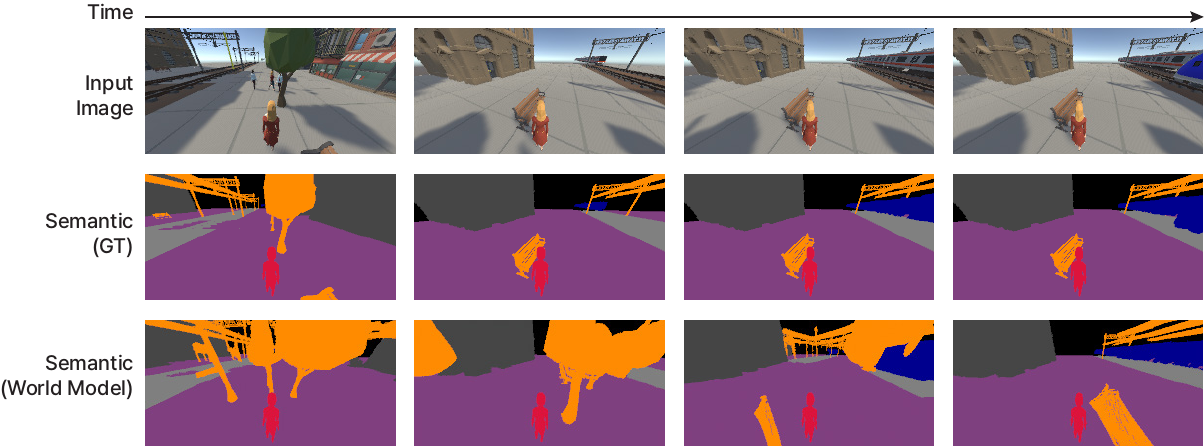}
\caption{\textbf{Segmentation image sequence from ETH\_Hotel.} Columns are ordered in time; rows are as in \autoref{fig:SemanticImages_Zara}.}
\label{fig:SemanticImages_Hotel}
\end{figure*}

\section{Additional Experiments for Motion Fidelity}
\label{app:motion_fidelity}

In addition to the closed-loop navigation metrics reported in the main paper,
we evaluate motion fidelity with Average Displacement Error (ADE) and Final
Displacement Error (FDE). These metrics compare generated trajectories against
reference human trajectories, where ADE measures the average point-wise
trajectory deviation and FDE measures the endpoint deviation. Lower values
indicate closer agreement with human motion.

\autoref{tab:motion_fidelity} reports results on held-out test trajectories from two scenes, Zara01 and Students003.
Our method achieves the lowest ADE in both scenes, indicating that the generated
trajectories remain closer to human demonstrations over the rollout horizon.
For FDE, our method obtains the best result on Zara01 and the second-best result
on Students003, where it remains close to the best baseline. These results
support the trajectory-fidelity aspect of our method: world-model-guided
preference refinement improves not only closed-loop navigation behavior, but
also the geometric similarity of generated motion to demonstrated human
trajectories.





\begin{table*}[t]
\centering
\small
\setlength{\tabcolsep}{6pt}
\renewcommand{\arraystretch}{1.05}
\caption{\textbf{Motion fidelity on held-out trajectories.}
ADE and FDE (in meters) compare generated trajectories against reference human trajectories.
Lower is better for both metrics; bold and underline mark the best and second-best results. Values are reported with 95\% confidence intervals.}
\label{tab:motion_fidelity}
\begin{tabular*}{\textwidth}{@{\extracolsep{\fill}}llcc@{}}
\toprule
Scene & Method & ADE $\downarrow$ & FDE $\downarrow$ \\
\midrule

\multirow{4}{*}{Zara01}
& GBM \citep{dutra2017gradient}
& $0.3057 \pm 0.0345$ & $0.5203 \pm 0.0509$ \\
& CCP \citep{panayiotou2022ccp}
& $\underline{0.2406 \pm 0.0005}$ & $\underline{0.4524 \pm 0.0005}$ \\
& CEDRL \citep{panayiotou2025cedrl}
& $0.3629 \pm 0.0071$ & $0.6817 \pm 0.0134$ \\
& \textbf{Ours}
& $\mathbf{0.2064 \pm 0.0069}$ & $\mathbf{0.3897 \pm 0.0141}$ \\
\midrule

\multirow{4}{*}{Students003}
& GBM \citep{dutra2017gradient}
& $0.2147 \pm 0.0026$ & $0.4194 \pm 0.0056$ \\
& CCP \citep{panayiotou2022ccp}
& $\underline{0.2053 \pm 0.0031}$ & $\mathbf{0.3873 \pm 0.0060}$ \\
& CEDRL \citep{panayiotou2025cedrl}
& $0.2338 \pm 0.0030$ & $0.4382 \pm 0.0058$ \\
& \textbf{Ours}
& $\mathbf{0.1994 \pm 0.0026}$ & $\underline{0.3948 \pm 0.0052}$ \\

\bottomrule
\end{tabular*}
\end{table*}

\section{User Study}
\label{app:user_study}

\begin{figure}[t]
    \centering
    \includegraphics[width=0.85\linewidth]{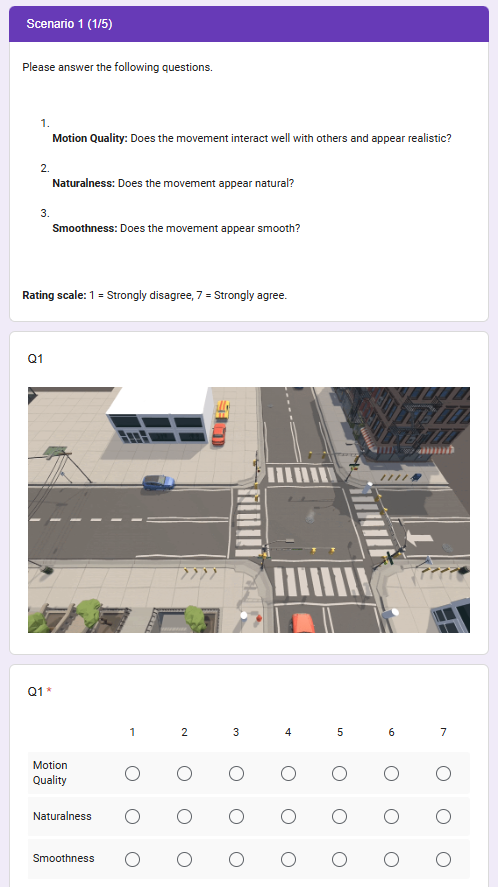}
        \caption{\textbf{Screenshot of the Google Form used for the user study.} The study was administered through Google Forms. Participants watched anonymized rollout videos and answered three 7-point Likert-scale questions evaluating motion quality, naturalness, and smoothness.}
    \label{fig:userstudyform}
\end{figure}

We conducted a user study to evaluate the perceived quality of the generated crowd navigation behaviors.
The study was administered through Google Forms, and a screenshot of the actual survey interface is
shown in \autoref{fig:userstudyform}.
A total of 20 participants took part in the study, with a balanced gender distribution
(10 male and 10 female; age $27.50 \pm 6.19$ years).
The study was approved by the Korea University Institutional Review Board (KUIRB-2026-0115-01), and all participants provided informed consent before participating.
Each participant received USD~\$7 as compensation for completing the study.

Each participant evaluated the results of four methods: GBM, CCP, CEDRL, and Ours.
We prepared 40 evaluation scenarios in total, consisting of 10 scenarios for each method.
The scenarios were evenly sampled from five evaluation scenes:
Zara01, Zara01\_Group, Students003, ETH\_Intersection, and ETH\_Hallway.
For each scenario, participants watched an anonymized rollout video and rated the resulting motion
using three 7-point Likert-scale questions: motion quality, naturalness, and smoothness.
Higher scores indicate better perceived motion quality.
The method identities were hidden from participants, and the presentation order was randomized to
reduce ordering bias.


For statistical analysis, we first averaged scores at the participant level for each method and each question,
rather than treating individual scenario ratings as independent samples.
This avoids pseudo-replication and respects the repeated-measures structure of the study.
Since all participants rated all methods and the scores are ordinal Likert ratings,
we used the Friedman test to assess the overall effect of method.
For post-hoc comparisons, we used paired Wilcoxon signed-rank tests comparing our method against each baseline,
with Holm correction for multiple comparisons.
We report statistical significance at two levels, $p<.05$ and $p<.01$.

\begin{figure}[t]
    \centering
    \includegraphics[width=\linewidth]{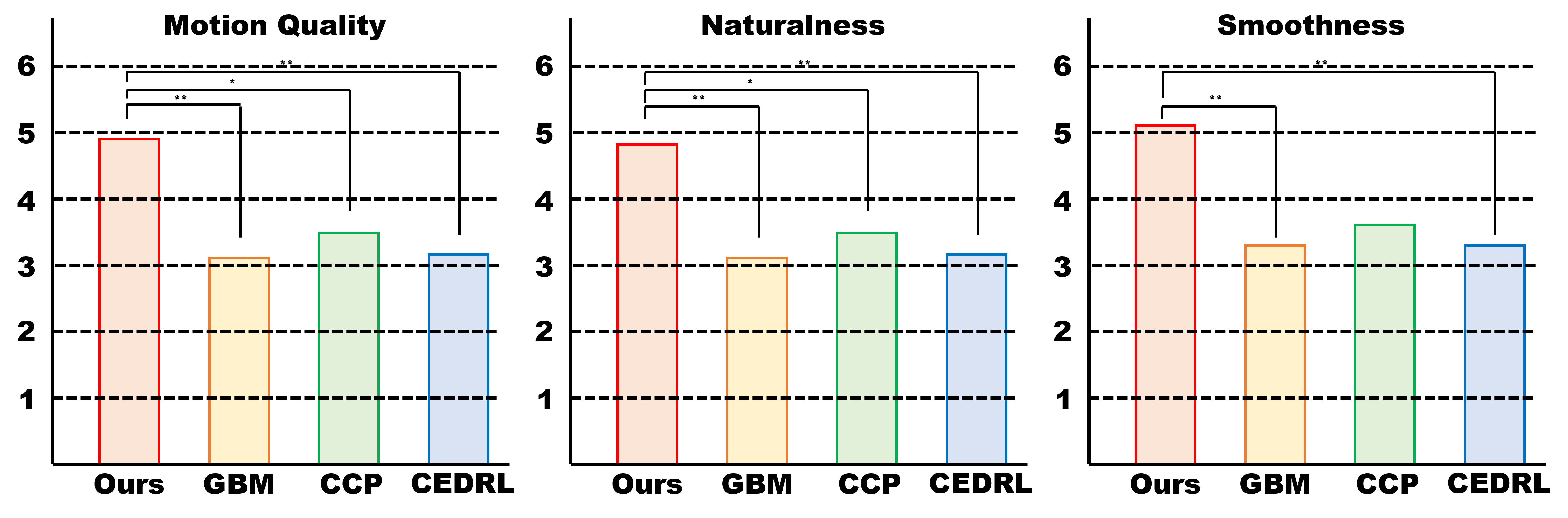}
    \caption{\textbf{User study results for motion quality, naturalness, and smoothness.}
    Bars show participant-level mean ratings. Brackets mark Holm-corrected paired Wilcoxon signed-rank tests comparing our method against each baseline (* $p<.05$, ** $p<.01$).}
    \label{fig:userstudy}
\end{figure}

As shown in \autoref{fig:userstudy}, our method received the highest mean score across all perceptual criteria.
Our method achieved an overall score of $4.94 \pm 0.44$, compared to
$3.19 \pm 0.34$ for GBM, $3.54 \pm 0.34$ for CCP, and $3.14 \pm 0.34$ for CEDRL.
For the individual criteria, our method obtained $4.91 \pm 0.44$ in motion quality,
$4.88 \pm 0.47$ in naturalness, and $5.05 \pm 0.44$ in smoothness.
The baselines received lower scores across the three criteria:
GBM achieved $3.14 \pm 0.32$, $3.13 \pm 0.35$, and $3.32 \pm 0.38$;
CCP achieved $3.48 \pm 0.32$, $3.45 \pm 0.35$, and $3.68 \pm 0.36$;
and CEDRL achieved $3.04 \pm 0.33$, $3.07 \pm 0.35$, and $3.31 \pm 0.36$
for motion quality, naturalness, and smoothness, respectively.
These results suggest that participants perceived our generated navigation behaviors as more natural,
higher-quality, and smoother than those of the baseline methods.

The Friedman test showed a significant overall effect of method for the aggregated score
($\chi^2(3)=20.84$, $p<.01$).
The effect was also significant for each individual criterion:
motion quality ($\chi^2(3)=21.28$, $p<.01$),
naturalness ($\chi^2(3)=16.18$, $p<.01$), and
smoothness ($\chi^2(3)=14.36$, $p<.01$).
These results indicate that the perceived quality differences among methods are statistically reliable
across all evaluation questions.

Post-hoc comparisons further show that our method was preferred over the baselines.
For the overall score, our method significantly outperformed GBM and CEDRL at the $p<.01$ level,
and CCP at the $p<.05$ level.
For motion quality, our method significantly outperformed GBM and CEDRL at the $p<.01$ level,
and CCP at the $p<.05$ level.
For naturalness, our method also significantly outperformed GBM and CEDRL at the $p<.01$ level,
and CCP at the $p<.05$ level.
For smoothness, our method significantly outperformed GBM and CEDRL at the $p<.01$ level,
while the difference against CCP was not statistically significant after correction.

Overall, the user study demonstrates that our method produces crowd navigation behaviors
that are perceived as higher-quality and more natural than those of existing baselines.
The improvements are consistent across motion quality and naturalness, where our method significantly
outperforms all three baselines.
For smoothness, our method still obtains the highest mean score, but its difference from CCP does not
reach statistical significance after correction.
This suggests that CCP can produce comparably smooth motion in some cases, while our method provides
stronger overall perceptual quality by better balancing naturalness, smoothness, and human-like
scene-compliant navigation.

\section{Qualitative Visualization of Segmentation Image Sequences}
\label{app:segmentation_sequences}

In addition to the quantitative semantic results in \autoref{app:world_model_performance}, we visualize segmentation image sequences predicted by the Semantic Transition World Model. 
These examples are used to inspect whether the model preserves navigation-relevant semantic structure over time. 
The goal is not photorealistic image generation, but temporally consistent prediction of semantic regions such as walkable space, road regions, buildings, obstacles, and vehicles.

\begin{figure*}[t]
\centering
\includegraphics[width=\textwidth]{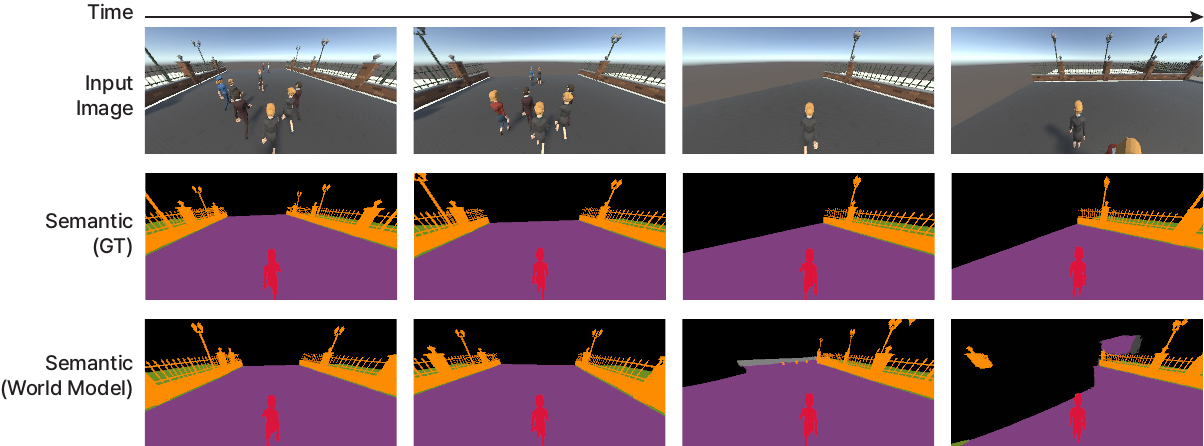}
\caption{\textbf{Segmentation image sequence from ETH.} Columns are ordered in time; rows are as in \autoref{fig:SemanticImages_Zara}.}
\label{fig:SemanticImages_ETH}
\end{figure*}

As shown in Figures \ref{fig:SemanticImages_Zara},  \ref{fig:SemanticImages_Hotel} and \ref{fig:SemanticImages_ETH}, the predicted segmentation sequences remain visually coherent across consecutive frames. 
Walkable regions largely preserve their spatial extent and the overall scene layout changes smoothly over time, although thin obstacles are sometimes over-extended (e.g., the orange regions in \autoref{fig:SemanticImages_Hotel}). 
This temporal consistency is important because the semantic prediction is used during policy refinement to evaluate whether each candidate anchor leads to a valid or invalid region.

The qualitative visualizations are consistent with the semantic prediction performance reported in \autoref{app:world_model_performance}. 
The model achieves high pixel accuracy, mIoU, Dice/F1, and frequency-weighted IoU on the test split, and the visualized sequences further show that these predictions remain stable over time. 
Thus, the Semantic Transition World Model provides a reliable navigation-oriented representation for semantic candidate scoring.

\begin{figure*}[t]
\centering
\includegraphics[width=\textwidth]{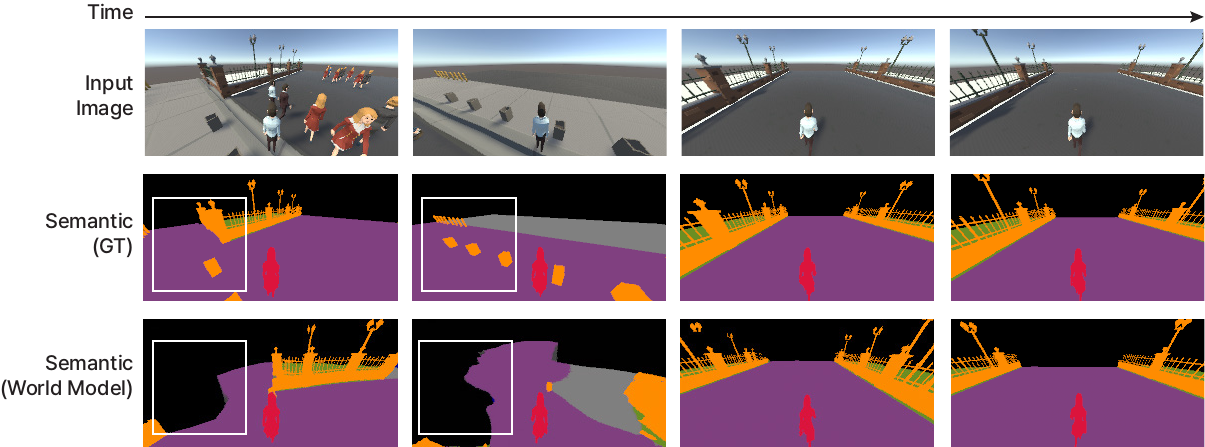}
\caption{\textbf{Failure case of a segmentation image sequence from ETH.} Rows are as in \autoref{fig:SemanticImages_Zara}. White boxes mark regions where the predicted segmentation deviates from the ground truth and loses temporal consistency.}
\label{fig:SemanticImages_ETH_Fail}
\end{figure*}

\section{Failure Cases and Limitations}
\label{app:failure_cases_limitations}

This appendix summarizes the main failure cases and limitations of the current method. 
Although the proposed policy benefits from the State Transition World Model and the Semantic Transition World Model during refinement, the world-model predictions are not always temporally perfect.

As shown in \autoref{fig:SemanticImages_ETH_Fail}, the Semantic Transition World Model can fail in complex ETH scenes. 
While the model captures the overall layout of the scene, some semantic regions change unnaturally across time or are not reconstructed consistently. 
For example, parts of the sidewalk, road boundary, obstacle region, or agent area may shift or disappear between consecutive predicted frames. 
This type of error can affect the semantic-validity score because the policy refinement relies on the predicted segmentation map to judge whether a candidate anchor moves toward a valid region.

Another limitation is that the current world models predict short-horizon consequences. 
This is sufficient for comparing immediate candidate anchors, but it may not capture long-range interaction patterns such as delayed yielding, dense crowd negotiation, or group reformation over several seconds. 
Thus, some actions that appear safe in the one-step prediction may still lead to undesirable behavior later in the rollout.

The policy also uses a fixed anchor vocabulary with $K=100$. 
This makes action selection efficient, but rare or unusual motions may be approximated imperfectly when they are not well covered by the anchor set. 
In addition, the semantic score depends on predefined segmentation labels and rule-based footprint thresholds, so transferring the method to new environments may require adjusting the semantic class mapping or validity rules.

Finally, all closed-loop experiments are conducted in reconstructed Unity scenes. 
Although the world models are removed at inference time and the deployed policy is efficient, real-world deployment would require additional validation under sensor noise, localization error, actuation delay, and stricter safety constraints. 
Therefore, the current results should be interpreted as evidence that world-model-guided refinement improves consequence-aware navigation, rather than as a complete real-robot safety guarantee.

\section{Runtime and Compute}
\label{app:runtime_compute_memory}

This appendix summarizes the compute and runtime cost of the world models used during policy refinement. 
The key property of our framework is that the State Transition World Model and the Semantic Transition World Model are used only during training-time candidate scoring. 
At inference time, the deployed policy directly selects an anchor from the current observation, and no world-model prediction or counterfactual scoring is performed. \autoref{tab:wm_compute_runtime} summarizes the compute and runtime of each component.

\begin{table}[t]
\centering
\caption{\textbf{World model and VLA compute and runtime.}
Runtime is measured per candidate for the world models and per decision for the deployed policy.}
\label{tab:wm_compute_runtime}
\small
\setlength{\tabcolsep}{0pt}
\begin{tabular*}{\linewidth}{@{\extracolsep{\fill}}lcc@{}}
\toprule
Component & Compute & Runtime \\
\midrule
Semantic Transition World Model & 31.1 TFLOPs & 0.115 s \\
State Transition World Model & 6.8 GFLOPs & 1.904 ms \\
Deployed policy (Qwen3-VL-2B + LoRA, anchor classification) & 23.6 TFLOPs & 65 ms \\
\bottomrule
\end{tabular*}
\end{table}

The Semantic Transition World Model is the dominant source of training-time compute. 
This is expected because it predicts a dense segmentation map through a denoising-based semantic transition model. 
In contrast, the State Transition World Model is lightweight because it predicts low-dimensional relative positions and velocities of nearby pedestrians and dynamic objects.

These costs are paid only when generating preference supervision during refinement. 
During deployment, the world models are removed and the policy performs a single anchor-selection step. 
Therefore, the additional computation introduced by consequence prediction improves training-time supervision without increasing test-time inference latency.

\section{Prompt Template}
\label{app:prompt_template}

The VLA policy receives a third-person RGB observation together with structured local navigation states and predicts one anchor index from the discrete anchor vocabulary. 
To make the input format consistent across supervised training, GRPO refinement, and inference, we use a fixed anchor-selection prompt template (\autoref{table:AnchorSelectionPrompt}). 
The prompt describes the coordinate convention, the ego state, the goal location, nearby pedestrians, nearby dynamic objects, and the available anchor-index range. 
All coordinates are expressed in the ego-centered local frame, where the controlled agent is located at $(0,0)$. 
Nearby pedestrians are listed within 7 meters, with same-group agents placed first and then non-group agents, each ordered from nearest to farthest. 
Nearby dynamic objects are listed within 15 meters and are also ordered by distance.

The output space is constrained to a single integer. 
This prevents the model from producing free-form language and makes the VLA policy directly compatible with the anchor classifier used in the rest of the method.

\begin{table*}[t]
\centering
\caption{\textbf{Anchor-selection prompt template.}
The prompt asks the policy to choose the safest and most natural anchor for crowd simulation based on the current image, ego state, goal, nearby agents, and nearby dynamic objects.}
\label{table:AnchorSelectionPrompt}
\begin{tcolorbox}[
    colback=gray!5,
    colframe=blue!40,
    title={Prompt Template},
    width=0.98\textwidth,
    boxrule=0.5pt,
    arc=1mm,
    left=2mm,
    right=2mm,
    top=1mm,
    bottom=1mm
]
\small
\textbf{Question:} The current image is your third-person view.\\
All coordinates below are relative to you, with your current position fixed at $(0, 0)$.\\
Your own velocity values are local motion values, and each nearby entity velocity is written as the other entity's velocity minus your current velocity.\\
Based on the scene and the state below, choose the anchor that is safest, goal-directed, collision-aware, socially compliant, smooth, and natural for crowd simulation.\\[0.5em]

Current local velocity: \texttt{\{velocity\}}.\\
Previous local velocity: \texttt{\{prev\_velocity\}}.\\
Goal position: \texttt{\{goal\_pos\}}.\\[0.5em]

Nearby agents within 7 meters 
(same-group agents first, then non-group agents; each part ordered from nearest to farthest):\\
\texttt{\{nearby\_agents\_text\}}\\[0.5em]

Nearby dynamic objects within 15 meters 
(ordered from nearest to farthest):\\
\texttt{\{nearby\_dynamic\_text\}}\\[0.5em]

Options: 0 to \texttt{\{num\_classes - 1\}}.\\
Answer with only one integer from 0 to \texttt{\{num\_classes - 1\}}.
\end{tcolorbox}
\end{table*}

\section{Additional Qualitative Results}
\label{app:additional_qualitative}

In this section, we provide additional qualitative results that complement the quantitative evaluation in the main paper. While scalar metrics summarize overall performance, they do not fully reveal how the agent interacts with surrounding pedestrians, vehicles, group members, and semantic scene regions over time. We therefore visualize representative closed-loop rollouts to better illustrate the behavioral differences between the baselines, ablated variants, and our full method.

\begin{figure*}
\centering
\includegraphics[width=\textwidth]{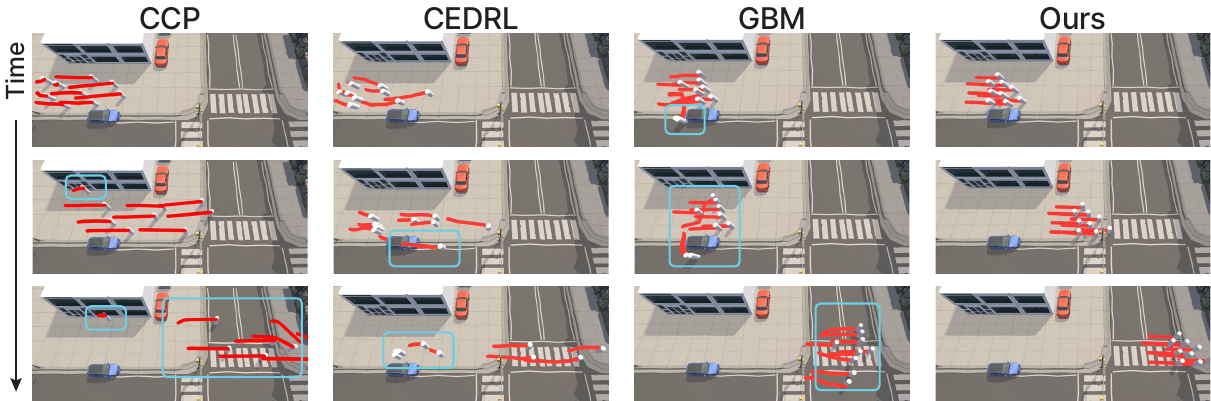}
\caption{\textbf{Additional qualitative comparison on group-coherent navigation.} We compare CCP, CEDRL, GBM, and our method in a representative group-navigation scene. Each row shows a later timestep in the same rollout, and trajectory traces indicate recent motion over the previous 60 frames. The highlighted regions show representative failure modes of the baseline methods, including group separation, collision with a vehicle, and excessive group spread that causes group members to move into road regions. By contrast, our method maintains a more compact and coherent relation among group members while avoiding vehicles and road regions, producing trajectories that better preserve group-level motion and scene compliance.}
\label{fig:GroupCoherence}
\end{figure*}

\autoref{fig:GroupCoherence} shows qualitative examples for group-coherent navigation. The goal of this setting is not only to reach the target while avoiding nearby pedestrians, but also to preserve a socially plausible relation with assigned group members. An action that is feasible for an individual agent can still be undesirable if it separates the agent from its group, produces unsafe interactions with vehicles, or stretches the group formation toward road regions.

Across the shown examples, the baseline methods often generate trajectories that are individually plausible but less consistent at the group level. In some cases, the controlled agent drifts away from its assigned group. In others, it fails to anticipate nearby vehicle motion, resulting in unsafe interactions. We also observe cases where the distance among group members becomes excessively large, causing the group formation to expand into semantically inappropriate road regions. These examples suggest that the baselines do not consistently balance goal-directed navigation, interaction safety, group coherence, and scene compliance.

By contrast, our method keeps the controlled agent closer to the group flow while still making progress toward the goal. The resulting trajectories remain more compact, avoid unsafe interactions with vehicles, and reduce the tendency of group members to spread into road regions. These qualitative results support the GD improvements reported in the main paper and show that the proposed consequence-based preference refinement improves not only numerical group-distance error, but also visually coherent and semantically safer group behavior.

\begin{figure*}
\centering
\includegraphics[width=\textwidth]{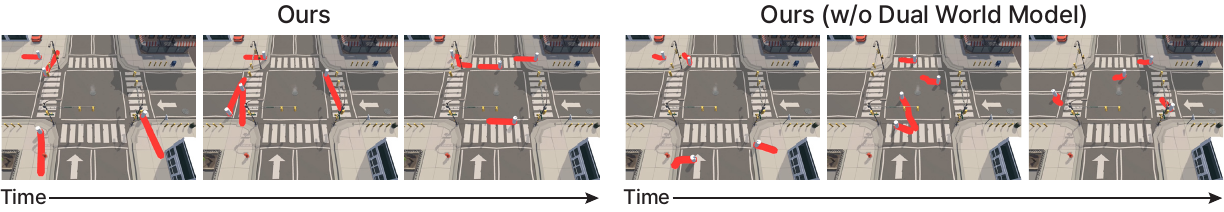}
\caption{\textbf{Qualitative ablation on the dual world models.} We compare our full method with the variant without the dual world models in a representative closed-loop navigation scene; time runs from left to right. Without the dual world models, the agent shows typical failure cases such as less anticipatory interaction behavior, unstable trajectory adjustment, or reduced compliance with scene semantics. In contrast, the full model, refined with world-model-predicted consequences, produces more stable, safe, and scene-aware trajectories.}
\label{fig:Ablation_WorldModel}
\end{figure*}

\autoref{fig:Ablation_WorldModel} further visualizes the effect of the dual world model. Without this component, the agent tends to rely more heavily on short-term local decisions, which can make the resulting behavior less stable in complex scenes. In particular, the ablated model may react too late to surrounding agents or dynamic objects, choose trajectories that are less compatible with the scene layout, or produce motion that appears less smooth over time.

The full method alleviates these issues by explicitly considering the predicted consequences of candidate actions during preference refinement. This allows the policy to better distinguish actions that are locally feasible from actions that remain safe and semantically appropriate over longer closed-loop rollouts. As a result, the full model produces trajectories that are more consistent with pedestrian motion, avoid unsafe interactions, and better respect scene constraints.

Together, these additional qualitative examples demonstrate that the proposed framework improves behavior beyond what is captured by individual metrics. The group-coherence examples show improved preservation of socially plausible group structure, while the ablation examples highlight the importance of the dual world model for anticipatory and scene-compliant navigation.

\begin{figure*}
\centering
\includegraphics[width=\textwidth]{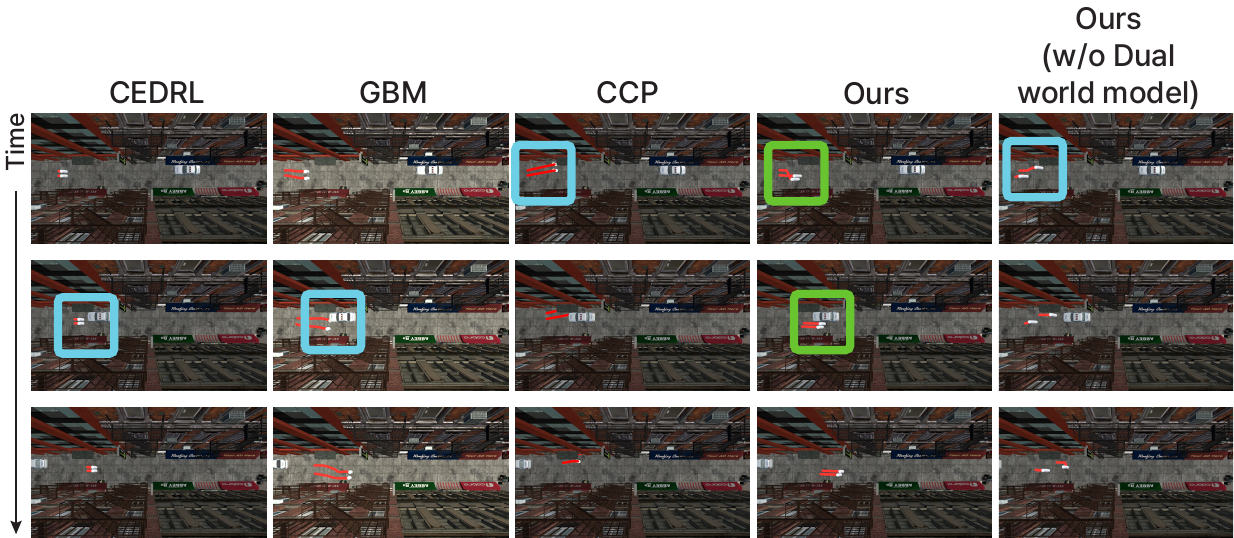}
\caption{\textbf{Narrow-alley interaction with an approaching vehicle.} Blue highlights indicate collision-prone trajectories produced by existing methods and the ablated variant, while green highlights show the collision-free trajectory produced by our method.}
\label{fig:Alley}
\end{figure*}

\autoref{fig:Alley} further shows that the proposed method can handle constrained interactions with dynamic vehicles. 
In this narrow-alley scenario, the baselines and the ablated variant often choose actions that appear locally plausible but enter the vehicle's future path. 
Our method instead anticipates this occupancy conflict and selects a safer side region, resulting in collision-free navigation through the alley.


\section{Comparison with Simpler Pipelines and an Online World-Model Planner}
\label{app:extra_baselines}

\autoref{tab:extra_baselines_full} reports the per-scene results summarized in \autoref{tab:extra_baselines}. Each trajectory predictor is combined with an MPC planner following~\cite{poddar2023crowd}, Gen-Safe-Nav~\cite{yao2025towards} integrates adaptive conformal inference into constrained reinforcement learning, and OnlineWM runs both world models at every decision. Our method achieves higher goal success and lower violation rates than all prediction-plus-planning pipelines and Gen-Safe-Nav in every scene, and stays close to OnlineWM across scenes.

\begin{table*}[t]
\centering
\small
\setlength{\tabcolsep}{4pt}
\renewcommand{\arraystretch}{1.05}
\caption{\textbf{Per-scene comparison with simpler pipelines and an online world-model planner.}
Trajectory predictors are combined with an MPC planner~\citep{poddar2023crowd}. Higher is better for GSR, while lower is better for SRVR and GD. Bold marks the best result in each scene among methods other than OnlineWM.}
\label{tab:extra_baselines_full}
\begin{tabular*}{\textwidth}{@{\extracolsep{\fill}}llccc@{}}
\toprule
Scene & Method & GSR $\uparrow$ & SRVR $\downarrow$ & GD $\downarrow$ \\
\midrule

\multirow{7}{*}{Zara01}
& OnlineWM & 78.9 & 5.7 & 0.6 \\
\cmidrule(l){2-5}
& Constant Velocity~\citep{scholler2020constant} & 51.8 & 15.1 & 0.7 \\
& Social-LSTM~\citep{alahi2016sociallstm} & 44.4 & 14.1 & 0.7 \\
& Social-GAN~\citep{gupta2018socialgan} & 44.4 & 15.0 & 1.0 \\
& Trajectron++~\citep{salzmann2020trajectron} & 48.2 & 13.8 & 0.8 \\
& Gen-Safe-Nav~\citep{yao2025towards} & 37.0 & 19.1 & 1.3 \\
& \textbf{Ours} & \textbf{72.2} & \textbf{7.0} & \textbf{0.6} \\
\midrule

\multirow{7}{*}{Zara01\_Group}
& OnlineWM & 77.8 & 0.8 & 1.7 \\
\cmidrule(l){2-5}
& Constant Velocity~\citep{scholler2020constant} & 25.9 & 18.1 & 6.7 \\
& Social-LSTM~\citep{alahi2016sociallstm} & 33.3 & 19.3 & 5.7 \\
& Social-GAN~\citep{gupta2018socialgan} & 37.0 & 14.8 & 6.3 \\
& Trajectron++~\citep{salzmann2020trajectron} & 25.9 & 19.0 & 5.9 \\
& Gen-Safe-Nav~\citep{yao2025towards} & 22.2 & 12.7 & 5.5 \\
& \textbf{Ours} & \textbf{77.8} & \textbf{1.2} & \textbf{2.1} \\
\midrule

\multirow{7}{*}{Students003}
& OnlineWM & 70.0 & 5.6 & 0.7 \\
\cmidrule(l){2-5}
& Constant Velocity~\citep{scholler2020constant} & 61.2 & 20.2 & 1.0 \\
& Social-LSTM~\citep{alahi2016sociallstm} & 54.0 & 14.6 & 1.1 \\
& Social-GAN~\citep{gupta2018socialgan} & 62.7 & 18.7 & 1.0 \\
& Trajectron++~\citep{salzmann2020trajectron} & 60.4 & 20.7 & 1.0 \\
& Gen-Safe-Nav~\citep{yao2025towards} & 32.7 & 11.6 & 1.3 \\
& \textbf{Ours} & \textbf{66.0} & \textbf{7.2} & \textbf{0.7} \\
\midrule

\multirow{7}{*}{ETH\_Intersection}
& OnlineWM & 88.9 & 3.6 & 0.8 \\
\cmidrule(l){2-5}
& Constant Velocity~\citep{scholler2020constant} & 66.7 & 11.5 & 1.5 \\
& Social-LSTM~\citep{alahi2016sociallstm} & 59.3 & 14.6 & 1.6 \\
& Social-GAN~\citep{gupta2018socialgan} & 59.3 & 12.5 & 1.5 \\
& Trajectron++~\citep{salzmann2020trajectron} & 66.7 & 12.8 & 1.5 \\
& Gen-Safe-Nav~\citep{yao2025towards} & 51.9 & 17.3 & 2.2 \\
& \textbf{Ours} & \textbf{83.3} & \textbf{3.9} & \textbf{0.9} \\
\midrule

\multirow{7}{*}{ETH\_Hallway}
& OnlineWM & 87.5 & 0.4 & 0.7 \\
\cmidrule(l){2-5}
& Constant Velocity~\citep{scholler2020constant} & 75.0 & 12.9 & 1.3 \\
& Social-LSTM~\citep{alahi2016sociallstm} & 68.6 & 12.3 & 1.4 \\
& Social-GAN~\citep{gupta2018socialgan} & 62.5 & 14.1 & 1.3 \\
& Trajectron++~\citep{salzmann2020trajectron} & 70.8 & 18.8 & 1.2 \\
& Gen-Safe-Nav~\citep{yao2025towards} & 70.8 & 9.9 & 1.4 \\
& \textbf{Ours} & \textbf{87.5} & \textbf{0.9} & \textbf{0.8} \\
\bottomrule
\end{tabular*}
\end{table*}

\section{Input Ablation}
\label{app:input_ablation}

To separate the contribution of the visual input from that of the hand-crafted state features, we train two variants: \emph{State only} removes the image, and \emph{Vision only} keeps the image and the goal but removes the other state features. As shown in \autoref{tab:input_ablation}, including vision generally improves GSR and SRVR, while the state features mainly help group coherence. Neither variant matches the full model, indicating that the two inputs are complementary.

\begin{table*}[t]
\centering
\small
\setlength{\tabcolsep}{4pt}
\renewcommand{\arraystretch}{1.05}
\caption{\textbf{Input ablation on five held-out evaluation scenarios.} State only removes the visual input; Vision only removes the hand-crafted state features except the goal. Higher is better for GSR, while lower is better for SRVR and GD.}
\label{tab:input_ablation}
\begin{tabular*}{\textwidth}{@{\extracolsep{\fill}}llccc@{}}
\toprule
Scene & Variant & GSR $\uparrow$ & SRVR $\downarrow$ & GD $\downarrow$ \\
\midrule

\multirow{3}{*}{Zara01}
& State only & 44.4 & 13.1 & \textbf{0.6} \\
& Vision only & 61.2 & 10.9 & 0.8 \\
& \textbf{Ours} & \textbf{72.2} & \textbf{7.0} & \textbf{0.6} \\
\midrule

\multirow{3}{*}{Zara01\_Group}
& State only & 44.4 & 12.3 & 2.2 \\
& Vision only & 44.4 & 6.4 & 2.3 \\
& \textbf{Ours} & \textbf{77.8} & \textbf{1.2} & \textbf{2.1} \\
\midrule

\multirow{3}{*}{Students003}
& State only & 50.0 & 12.9 & \textbf{0.7} \\
& Vision only & 54.0 & 8.7 & 0.8 \\
& \textbf{Ours} & \textbf{66.0} & \textbf{7.2} & \textbf{0.7} \\
\midrule

\multirow{3}{*}{ETH\_Intersection}
& State only & 55.6 & 10.3 & 1.1 \\
& Vision only & 66.7 & 10.0 & 2.1 \\
& \textbf{Ours} & \textbf{83.3} & \textbf{3.9} & \textbf{0.9} \\
\midrule

\multirow{3}{*}{ETH\_Hallway}
& State only & 62.5 & 1.6 & 1.0 \\
& Vision only & 68.8 & 1.1 & 1.6 \\
& \textbf{Ours} & \textbf{87.5} & \textbf{0.9} & \textbf{0.8} \\

\bottomrule
\end{tabular*}
\end{table*}

\section{Standard ETH/UCY Forecasting Benchmark}
\label{app:standard_benchmark}

To test whether our framework is tied to the Unity reconstructions, we retrain and evaluate it under the standard ETH/UCY forecasting protocol with leave-one-scene-out evaluation. Social-LSTM and Social-GAN numbers are taken from~\cite{gupta2018socialgan}, and Trajectron++ numbers are those obtained after correcting an evaluation bug in its original implementation~\cite{xu2023eqmotion}. As shown in \autoref{tab:standard_benchmark}, our policy performs comparably to Trajectron++. Our goal is not state-of-the-art forecasting accuracy; the result indicates that the framework remains competitive outside our simulation.

\begin{table*}[t]
\centering
\small
\setlength{\tabcolsep}{4pt}
\renewcommand{\arraystretch}{1.05}
\caption{\textbf{ETH/UCY forecasting benchmark (ADE/FDE in meters, lower is better).} Trajectron++ numbers are those after the evaluation bug fix~\cite{xu2023eqmotion}.}
\label{tab:standard_benchmark}
\begin{tabular*}{\textwidth}{@{\extracolsep{\fill}}lccc|ccc@{}}
\toprule
& \multicolumn{3}{c|}{Most-likely} & \multicolumn{3}{c}{Best-of-20} \\
Dataset & Social-LSTM~\citep{alahi2016sociallstm} & Trajectron++~\citep{salzmann2020trajectron} & Ours & Social-GAN~\citep{gupta2018socialgan} & Trajectron++~\citep{salzmann2020trajectron} & Ours \\
\midrule
ETH     & 1.09/2.35 & 1.02/2.00 & 1.05/2.08 & 0.81/1.52 & 0.67/1.18 & 0.65/1.08 \\
Hotel   & 0.79/1.76 & 0.33/0.62 & 0.37/0.68 & 0.72/1.61 & 0.18/0.28 & 0.25/0.30 \\
Univ    & 0.67/1.40 & 0.53/1.19 & 0.40/1.02 & 0.60/1.26 & 0.30/0.54 & 0.23/0.50 \\
Zara1   & 0.47/1.00 & 0.44/0.99 & 0.38/0.90 & 0.34/0.69 & 0.25/0.41 & 0.22/0.40 \\
Zara2   & 0.56/1.17 & 0.32/0.73 & 0.35/0.79 & 0.42/0.84 & 0.18/0.32 & 0.20/0.38 \\
\midrule
Average & 0.72/1.54 & 0.53/1.11 & 0.51/1.10 & 0.58/1.18 & 0.31/0.54 & 0.31/0.53 \\
\bottomrule
\end{tabular*}
\end{table*}

\clearpage 


\newpage
\input{checklist.tex}

\end{document}

%% file: checklist.tex
\section*{NeurIPS Paper Checklist}
\begin{enumerate}

\item {\bf Claims}
    \item[] Question: Do the main claims made in the abstract and introduction accurately reflect the paper's contributions and scope?
    \item[] Answer: \answerYes{} 
    \item[] Justification: The claims in the abstract and introduction are supported by the proposed method, the closed-loop evaluation, the ablation study, and the user study. We explicitly evaluate goal success, semantic/social violations, group coherence, motion fidelity, and perceived human-likeness.

\item {\bf Limitations}
    \item[] Question: Does the paper discuss the limitations of the work performed by the authors?
    \item[] Answer: \answerYes{} 
    \item[] Justification: We discuss limitations in Section~\ref{sec:limitations} and failure cases in Appendix~\ref{app:failure_cases_limitations}.

\item {\bf Theory assumptions and proofs}
    \item[] Question: For each theoretical result, does the paper provide the full set of assumptions and a complete (and correct) proof?
    \item[] Answer: \answerNA{} 
    \item[] Justification: The paper does not introduce formal theoretical results, theorems, or proofs.

    \item {\bf Experimental result reproducibility}
    \item[] Question: Does the paper fully disclose all the information needed to reproduce the main experimental results of the paper to the extent that it affects the main claims and/or conclusions of the paper (regardless of whether the code and data are provided or not)?
    \item[] Answer: \answerYes{} 
    \item[] Justification: We provide sufficient details to reproduce the method, including the observation format, anchor vocabulary construction, supervised training objective, world-model objectives, GRPO refinement procedure, score function, hyperparameters, evaluation scenes, and metrics.

\item {\bf Open access to data and code}
    \item[] Question: Does the paper provide open access to the data and code, with sufficient instructions to faithfully reproduce the main experimental results, as described in supplemental material?
    \item[] Answer: \answerYes{} 
    \item[] Justification: The project page is publicly available at \url{https://seoneun.github.io/-NeurIPS-2026-Seeing-Beyond-the-Next-Step/}. The reconstructed Unity scenes are built on commercially licensed assets that cannot be redistributed, so they are not released; the ETH/UCY trajectories are available from their original sources.

\item {\bf Experimental setting/details}
    \item[] Question: Does the paper specify all the training and test details (e.g., data splits, hyperparameters, how they were chosen, type of optimizer) necessary to understand the results?
    \item[] Answer: \answerYes{} 
    \item[] Justification: We describe the experimental setup, datasets, evaluation scenarios, baselines, metrics, training hyperparameters, and implementation details. We also provide additional appendix details for the score function, segmentation classes, world-model diagnostics, and runtime/compute cost.

\item {\bf Experiment statistical significance}
    \item[] Question: Does the paper report error bars suitably and correctly defined or other appropriate information about the statistical significance of the experiments?
    \item[] Answer: \answerYes{} 
    \item[] Justification: We report results over multiple runs and include uncertainty estimates where applicable. Motion fidelity results are reported with 95\% confidence intervals, and the closed-loop experiments are conducted across multiple evaluation scenarios and repeated runs.

\item {\bf Experiments compute resources}
    \item[] Question: For each experiment, does the paper provide sufficient information on the computer resources (type of compute workers, memory, time of execution) needed to reproduce the experiments?
    \item[] Answer: \answerYes{} 
    \item[] Justification: We report implementation and compute-related details, including model backbone, LoRA settings, training hyperparameters, denoising steps, world-model runtime, and compute cost. We also clarify that world models are used only during training-time refinement and are removed at inference.
    
\item {\bf Code of ethics}
    \item[] Question: Does the research conducted in the paper conform, in every respect, with the NeurIPS Code of Ethics \url{https://neurips.cc/public/EthicsGuidelines}?
    \item[] Answer: \answerYes{} 
    \item[] Justification: We have reviewed and followed the NeurIPS Code of Ethics.

\item {\bf Broader impacts}
    \item[] Question: Does the paper discuss both potential positive societal impacts and negative societal impacts of the work performed?
    \item[] Answer: \answerNo{} 
    \item[] Justification: The current version does not include a dedicated broader-impact discussion. The work focuses on human-like navigation in reconstructed simulation environments and does not make claims about direct deployment in real-world safety-critical systems.
    
\item {\bf Safeguards}
    \item[] Question: Does the paper describe safeguards that have been put in place for responsible release of data or models that have a high risk for misuse (e.g., pre-trained language models, image generators, or scraped datasets)?
    \item[] Answer: \answerNA{} 
    \item[] Justification: The paper does not release a high-risk foundation model, autonomous real-world control system, or other artifact that requires additional safeguards beyond the reported simulation-based evaluation.

\item {\bf Licenses for existing assets}
    \item[] Question: Are the creators or original owners of assets (e.g., code, data, models), used in the paper, properly credited and are the license and terms of use explicitly mentioned and properly respected?
    \item[] Answer: \answerYes{} 
    \item[] Justification: We cite the datasets, pretrained models, and baseline methods used in the experiments. We also respect licensing constraints for external assets and do not redistribute assets that cannot be publicly released.

\item {\bf New assets}
    \item[] Question: Are new assets introduced in the paper well documented and is the documentation provided alongside the assets?
    \item[] Answer: \answerNA{} 
    \item[] Justification: We do not release new third-party assets. The reconstructed Unity scenes rely on external assets whose licenses do not permit redistribution, so the assets themselves will not be released.

\item {\bf Crowdsourcing and research with human subjects}
    \item[] Question: For crowdsourcing experiments and research with human subjects, does the paper include the full text of instructions given to participants and screenshots, if applicable, as well as details about compensation (if any)? 
    \item[] Answer: \answerYes{} 
    \item[] Justification: We describe the user study protocol in Appendix~\ref{app:user_study}, including the number of participants, anonymized rollout videos, randomized presentation order, and 7-point Likert-scale questions for motion quality, naturalness, and smoothness. A screenshot of the survey interface is shown in Figure~\ref{fig:userstudyform}, and each participant received USD~\$7 as compensation.


\item {\bf Institutional review board (IRB) approvals or equivalent for research with human subjects}
    \item[] Question: Does the paper describe potential risks incurred by study participants, whether such risks were disclosed to the subjects, and whether Institutional Review Board (IRB) approvals (or an equivalent approval/review based on the requirements of your country or institution) were obtained?
    \item[] Answer: \answerYes{} 
    \item[] Justification: The user study was approved by the Korea University Institutional Review Board (KUIRB-2026-0115-01), as stated in Appendix~\ref{app:user_study}. The study involved only watching simulation videos and answering rating questions, posing minimal risk to participants.

\item {\bf Declaration of LLM usage}
    \item[] Question: Does the paper describe the usage of LLMs if it is an important, original, or non-standard component of the core methods in this research? Note that if the LLM is used only for writing, editing, or formatting purposes and does \emph{not} impact the core methodology, scientific rigor, or originality of the research, declaration is not required.
    \item[] Answer: \answerYes{} 
    \item[] Justification: The method uses a pretrained vision-language model as the policy backbone, which is adapted into a vision-language-action policy through supervised training and GRPO-based refinement. We clearly describe how this model is used as part of the proposed method.

\end{enumerate}